\documentclass[12pt,preprint]{aastex}
 
\newcommand{\IP}{ionization parameter}
\newcommand{\IPs}{ionization parameters}

\newcommand{\qals}{quasar absorption lines}

\newcommand{\etal}{et~al.\ }
\newcommand{\PVdblt}{{\rm P}\kern 0.1em{\sc v}~$\lambda\lambda 1117, 1128$}
\newcommand{\CaIIdblt}{{\rm Ca}\kern 0.1em{\sc ii}~$\lambda\lambda 3934, 3969$}
\newcommand{\AlIIIdblt}{{\rm Al}\kern 0.1em{\sc iii}~$\lambda\lambda 1854, 1862$}
\newcommand{\CIVdblt}{{\rm C}\kern 0.1em{\sc iv}~$\lambda\lambda 1548, 1551$}
\newcommand{\MgIIdblt}{{\rm Mg}\kern 0.1em{\sc ii}~$\lambda\lambda 2796, 2803$}
\newcommand{\NVdblt}{{\rm N}\kern 0.1em{\sc v}~$\lambda\lambda 1239, 1243$}  
\newcommand{\SVIdblt}{{\rm S}\kern 0.1em{\sc vi}~$\lambda\lambda 933, 944$} 
\newcommand{\OVIdblt}{{\rm O}\kern 0.1em{\sc vi}~$\lambda\lambda 1032, 1038$} 
\newcommand{\SiIIdblt}{{\rm Si}\kern 0.1em{\sc ii}~$\lambda\lambda 1190, 1193$} 
\newcommand{\SiIVdblt}{{\rm Si}\kern 0.1em{\sc iv}~$\lambda\lambda 1394, 1403$} 
\newcommand{\PV}{\hbox{{\rm P}\kern 0.1em{\sc v}}}
\newcommand{\AlI}{\hbox{{\rm Al}\kern 0.1em{\sc i}}}
\newcommand{\AlII}{\hbox{{\rm Al}\kern 0.1em{\sc ii}}}
\newcommand{\AlIII}{{\hbox{\rm Al}\kern 0.1em{\sc iii}}}
\newcommand{\CaII}{\hbox{{\rm Ca}\kern 0.1em{\sc ii}}}
\newcommand{\CII}{\hbox{{\rm C}\kern 0.1em{\sc ii}}}
\newcommand{\CIIe}{\hbox{{\rm C$^{\ast}$}\kern 0.1em{\sc ii}}}
\newcommand{\CIII}{\hbox{{\rm C}\kern 0.1em{\sc iii}}}
\newcommand{\CIV}{\hbox{{\rm C}\kern 0.1em{\sc iv}}}
\newcommand{\CV}{\hbox{{\rm C}\kern 0.1em{\sc v}}}
\newcommand{\HI}{\hbox{{\rm H}\kern 0.1em{\sc i}}}
\newcommand{\HII}{\hbox{{\rm H}\kern 0.1em{\sc ii}}}
\newcommand{\Lya}{\hbox{{\rm Ly}\kern 0.1em$\alpha$}}
\newcommand{\Lyb}{\hbox{{\rm Ly}\kern 0.1em$\beta$}}
\newcommand{\Lyg}{\hbox{{\rm Ly}\kern 0.1em$\gamma$}}
\newcommand{\Lyd}{\hbox{{\rm Ly}\kern 0.1em$\delta$}}
\newcommand{\Lye}{\hbox{{\rm Ly}\kern 0.1em$\epsilon$}}
\newcommand{\Lyphi}{\hbox{{\rm Ly}\kern 0.1em$\phi$}}
\newcommand{\Lyfive}{\hbox{{\rm Ly}\kern 0.1em$5$}}
\newcommand{\Lysix}{\hbox{{\rm Ly}\kern 0.1em$6$}}
\newcommand{\Lyseven}{\hbox{{\rm Ly}\kern 0.1em$7$}}
\newcommand{\Lyeight}{\hbox{{\rm Ly}\kern 0.1em$8$}}
\newcommand{\Lynine}{\hbox{{\rm Ly}\kern 0.1em$9$}}
\newcommand{\Lyten}{\hbox{{\rm Ly}\kern 0.1em$10$}}
\newcommand{\Lyeleven}{\hbox{{\rm Ly}\kern 0.1em$11$}}
\newcommand{\HeI}{\hbox{{\rm He}\kern 0.1em{\sc i}}}
\newcommand{\HeII}{\hbox{{\rm He}\kern 0.1em{\sc ii}}}
\newcommand{\FeI}{\hbox{{\rm Fe}\kern 0.1em{\sc i}}}
\newcommand{\FeII}{\hbox{{\rm Fe}\kern 0.1em{\sc ii}}}
\newcommand{\FeIII}{\hbox{{\rm Fe}\kern 0.1em{\sc iii}}}
\newcommand{\MnII}{\hbox{{\rm Mn}\kern 0.1em{\sc ii}}}
\newcommand{\MgI}{\hbox{{\rm Mg}\kern 0.1em{\sc i}}}
\newcommand{\MgII}{\hbox{{\rm Mg}\kern 0.1em{\sc ii}}}
\newcommand{\MgIII}{\hbox{{\rm Mg}\kern 0.1em{\sc iii}}}
\newcommand{\NI}{\hbox{{\rm N}\kern 0.1em{\sc i}}}
\newcommand{\NII}{\hbox{{\rm N}\kern 0.1em{\sc ii}}}
\newcommand{\NIII}{\hbox{{\rm N}\kern 0.1em{\sc iii}}}
\newcommand{\NV}{\hbox{{\rm N}\kern 0.1em{\sc v}}}
\newcommand{\OVI}{\hbox{{\rm O}\kern 0.1em{\sc vi}}}
\newcommand{\OI}{\hbox{{\rm O}\kern 0.1em{\sc i}}}
\newcommand{\OII}{\hbox{{\rm O}\kern 0.1em{\sc ii}}}
\newcommand{\OIV}{\hbox{{\rm O}\kern 0.1em{\sc iv}}}
\newcommand{\SI}{{\rm S}\kern 0.1em{\sc i}}
\newcommand{\SIV}{{\rm S}\kern 0.1em{\sc iv}}
\newcommand{\SVI}{{\rm S}\kern 0.1em{\sc vi}}
\newcommand{\SiI}{\hbox{{\rm Si}\kern 0.1em{\sc i}}}
\newcommand{\SiII}{\hbox{{\rm Si}\kern 0.1em{\sc ii}}}
\newcommand{\SiIII}{\hbox{{\rm Si}\kern 0.1em{\sc iii}}}
\newcommand{\SiIV}{\hbox{{\rm Si}\kern 0.1em{\sc iv}}}
\newcommand{\SII}{\hbox{{\rm S}\kern 0.1em{\sc ii}}}
\newcommand{\SIII}{\hbox{{\rm S}\kern 0.1em{\sc iii}}}
\newcommand{\NaI}{\hbox{{\rm Na}\kern 0.1em{\sc i}}}
\newcommand{\TiII}{\hbox{{\rm Ti}\kern 0.1em{\sc ii}}}
\newcommand{\kms}{\hbox{km~s$^{-1}$}}
\newcommand{\cmsq}{\hbox{cm$^{-2}$}}
\newcommand{\cc}{\hbox{cm$^{-3}$}}
 
\tighten
\begin{document}

\received{29 July 2004}
\accepted{23 December 2004}
\journalid{623}{10 April 2005}
\slugcomment{accepted by: {\it The Astrophysical Journal}}
 
\shortauthors{MASIERO ET~AL.}
\shorttitle{Models of Five Absorption Line Systems}


\title{Models of Five Absorption Line Systems Along the Line of Sight Toward PG0117+213\altaffilmark{1,2}}

\author{Joseph~R.~Masiero\altaffilmark{3}, Jane~C.~Charlton\altaffilmark{3}, Jie~Ding\altaffilmark{3},
Christopher~W.~Churchill\altaffilmark{4,5} and Glenn Kacprzak\altaffilmark{4}}

\altaffiltext{1}{Based in part on observations obtained at the
W.~M. Keck Observatory, which is operated as a scientific partnership
among Caltech, the University of California, and NASA. The Observatory
was made possible by the generous financial support of the W. M. Keck
Foundation.}
\altaffiltext{2}{Based in part on observations obtained with the
NASA/ESA {\it Hubble Space Telescope}, which is operated by the STScI
for the Association of Universities for Research in Astronomy, Inc.,
under NASA contract NAS 5-26555.}
\altaffiltext{3}{Department of Astronomy and Astrophysics, The Pennsylvania State University, University Park, PA 16802, {\it masiero, charlton, ding@astro.psu.edu}}
\altaffiltext{4}{Department of Astronomy, New Mexico State University
  1320 Frenger Mall, Las Cruces, New Mexico 88003-8001, {\it cwc, glennk@nmsu.edu}}
\altaffiltext{5}{Visiting Astronomer at the W.~M. Keck Observatory}

\begin{abstract}

We present our investigation into the physical conditions of the gas
in five intervening quasar absorption line systems along the line of
sight toward the quasar PG~$0117+213$, with redshifts of $z=0.57$,
$z=0.72$, $z=1.04$, $z=1.32$ and $z=1.34$.  Photoionization modeling
of {\it HST}, Keck I, and Palomar data, using the code Cloudy, is
employed to derive densities and metallicities of the multiple phases
of gas required to fit the absorption profile for each system.  We
discuss the implications of these models for galaxy evolution,
including the interpretation of ``{\CIV} deficiency'' and damped Lyman
alpha absorbers (DLAs), and the relationships between galaxy morphology,
galaxy luminosity, and absorption signature.
\newpage
\end{abstract}



\section{Introduction}
Quasar absorption line analysis is a powerful tool for studying the
properties of different types of gas in the universe, including gas
around quasars, disks and halos of intervening galaxies, and high
velocity clouds.  This paper focuses on the {\MgII} absorption line
systems produced by five intervening galaxies along the line of sight
toward the quasar PG~$0117+213$.  The most difficult part of analyzing
{\qals} is connecting the velocity space absorption profiles that are
observed in the quasar's spectrum with the physical properties of the
gas that is creating the profile.  In particular, trying to associate
intervening absorption systems with particular types of structures in
galaxies has proven to be difficult.  Two methods for making this
association are imaging the field around the quasar and looking for
galaxies at the same redshifts as the absorption systems, and
indirectly comparing properties determined from the absorption
profiles with physical properties and processes that are known to
occur in galaxy halos and disks.  There are difficulties with each,
however.  Imaging a galaxy becomes practically impossible for large
redshifts, dwarf galaxy hosts, or galaxies at small impact parameters
from the quasar.  On the other hand, velocity profiles can be very
complicated, involving rotation, thermal components, gaseous
expulsion, collisions, unresolved features, and unrelated absorption.

The {\MgIIdblt} doublet is a key identification tool for absorption
systems, due to its strength, $2:1$ doublet ratio, and location in the
optical for moderate redshifts ($0.3<z<2.4$).  {\MgII} absorbers can
be divided into two different types: weak, with equivalent widths
$<0.3~$\AA, and strong, with equivalent widths $\ge0.3~$\AA.  The
division of $0.3~$\AA~ was historically chosen as an artifact of
observational sensitivity \citep{ss92}, however it also may have some
significance as the approximate transition point in the neutral
hydrogen between optically thin and optically thick \citep{weak2}.
Weak {\MgII} systems can be further subdivided into two main
categories. Single cloud weak absorbers are isolated, narrow clouds with
Doppler parameters $2<b<8$~{\kms} \citep{weak1}.  Multiple cloud weak
absorbers are a collection of single, weak {\MgII} clouds within a
narrow, related velocity range (e.g. $300$~{\kms}) with a total
equivalent width of $<0.3~$\AA.  Typically, strong {\MgII} absorbers
are saturated, and weak ones are unsaturated, however there are
important exceptions to this generalization.  One example of such a
case would be a multiple cloud system with a total equivalent width
greater than $0.3~$\AA~(and therefore classified as a strong {\MgII}
system), which in actuality is a collection of unsaturated weak
{\MgII} clouds. Also, it should be noted that an unresolved weak
{\MgII} cloud can be saturated while not being completely black in the
core.  This is a very important phenomenon to be aware of when
modeling very narrow weak {\MgII} clouds, as it can strongly affect
the modeling results \citep{ding1634}.

Single cloud weak {\MgII} absorbers have properties consistent with
small pockets or sheets of low ionization gas associated with
highly-ionized extended gas structures, but most are not known to be
associated with any intervening galaxy \citep{weak2}.  One multiple
cloud weak {\MgII} absorber has been hypothesized to be a superwind
coming off of one or a pair of dwarf galaxies \citep{zonak}.  Others
are thought to be collections of weak clouds all belonging to the same
host galaxy as extensions of the strong {\MgII} absorber population.

In the past, high resolution coverage of multiple ionic transitions
enabled the modeling of physical conditions of quasar absorption line
systems at low redshift \citep{sembach95,tripp96,tripp97,sembach99,
chen00}.  This paper is not the first to analyze intermediate
redshift, $z\sim1$, {\MgII} systems along the line of sight toward a
quasar, however few other absorption systems at that redshift have
been analyzed to the level of detail we present here.  Previous
detailed studies that have used the same modeling method include
\citet{ding1206}, who modeled what they believe to be a double galaxy
at redshift $z=0.93$ along the line of sight toward PG~$1206+459$,
which had drastically different high ionization structure for each of
the galaxies, and
\citet{ding1634}, who considered an absorption system at $z=0.99$
toward PG~$1634+706$ which was deficient in {\CIV}, potentially due to
a low metallicity or very high ionization broad {\HI} absorption
phase.  Several weak {\MgII} absorbers, single and multiple cloud,
have been modeled in detail
\citep{weak1634,zonak}.  We have also recently completed a similar
study of six {\MgII} absorption systems along the PG~$1241+176$,
PG~$1248+401$, and PG~$1317+274$ lines of sight \citep{ding04}.  Each new
system studied in such detail has added new insights into the mix of
processes that determine gas structure formation.

The number of detailed studies at $z\sim1$ has been limited because both
high resolution and coverage of many different absorption species are required
to accurately constrain a model.  PG~$0117+213$ is ideal for a study
like this because it is bright in both the optical and UV, and it has
five intervening {\MgII} absorbers in one high resolution spectrum.
These systems were discussed in an earlier paper by \citet{archiveI},
however they did not have the benefit of the high resolution UV
coverage of many important transitions by the Space Telescope Imaging
Spectrograph onboard the {\it Hubble Space Telescope} ({\it
HST}/STIS).

In this paper, we present photoionization and collisional ionization
models of the subcomponents of the five intervening {\MgII} systems,
also known as clouds, which were made to determine the physical
properties of the gas, such as {\IP} and metallicity.  Our model
clouds were ionized by the Haardt-Madau extragalactic background
radiation \citep{hm}.  The values for the {\IPs} of the clouds are
used to separate them into distinct regimes called phases.  These
phases, characterized by the density of the gas, are broadly divided
by the different types of ionization they exhibit: high, low, or
very-low.  Many absorption line systems show multiple phases that
overlap in velocity space, but have very distinct kinematic profiles.
When many absorption species are available as constraints, applying
models with overlapping phases of ionization to these multi-phase
systems allows for very detailed models of the absorption lines to be
created.

We describe our data and processing techniques in \S~\ref{d&p}.  Next,
we describe in depth our method for modeling {\qals} in
\S~\ref{mtech}.  We then present our modeling results and a discussion
of their significance in \S~\ref{results}.  Note that in this section,
we discuss each of the five systems in separate subsections.  We
summarize our results in \S~\ref{sum}, and state our conclusions in
\S~\ref{conc}.

\section{Data and Processing}
\label{d&p}

\subsection{The Data}
\label{data}
Spectra from the {\it HST}, the Keck I telescope, and the Palomar
$200~$inch telescope were used to constrain our models.  The Keck I
High Resolution Spectrometer (HIRES, \cite{vogt94}) spectrum covers a
wavelength range of $4317.7~$\AA~ to $6775.1~$\AA, with a resolution
of $45000$ and a signal-to-noise of around $30$ per pixel.  This
spectrum was reduced in the standard way using IRAF, as described in
\citet{archiveI}.  The {\it HST}/STIS spectrum was obtained using the
E230M grating with a $0.2"$ slit width, and a central wavelength of
$2707~$\AA.  The spectrum covers a wavelength range of $2303$ to
$3111~$\AA, with a resolution of $30000$.  This spectrum was reduced
with the standard STIS pipeline \citep{stis1}.  The overlapping
sections of the spectrum were co-added.  The combined spectrum has a
signal-to-noise range of $4-9$ per pixel, based on the standard
deviation of the data points.

A {\it HST} Faint Object Spectrograph (FOS) spectrum, obtained with
the G190H grating, covers a wavelength range of $1574~$\AA~ to
$2331~$\AA, and has a resolution of $1300$.  This spectrum was reduced
as part of the QSO Absorption Line Key Project \citep{cat1,cat2,cat3}.
This spectrum was obtained using the spectropolarimetry mode, with a
$1.0''$ aperture \citep{koratkar}.  Only the {\Lya} line from the
$z=0.5764$ damped {\Lya} absorber (DLA) was detected in this
relatively noisy spectrum.  Because of this, we use it primarily as a
constraint on the Lyman limit breaks of our systems.  The ground based
spectrum covering the {\CIV} doublet for the $z=1.3250$ and $z=1.3430$
systems was taken from \citet{ss92}, who used the Palomar $200~$inch
telescope.  This spectrum has a resolution of $860$.

To fit the STIS spectrum, the echelle orders were combined into one
large file, and then broken up into a few large regions.  The
unabsorbed sections of continuum were chosen for each region, and
variable order polynomials were fit to the continuum.  The spectrum
was then normalized with these polynomials in order to facilitate a
search for absorption features.  The HIRES spectrum was fit with
Legendre polynomials as described in Churchill {\etal} (2000a) (and
references therein).

Because the G190H FOS data covered the three Lyman limit breaks almost
exclusively, there were very few points that could be considered
continuum, and so the continuum was generally assumed to be flat in
the region redward of the $z=1.3430$ Lyman limit break, and
extrapolated over the Lyman limit breaks.

The data from all the instruments that is relevant to constraining
each of the five systems is presented in Figures
\ref{fig134}-\ref{fig057}.  Tables \ref{ew134}-\ref{ew057} present the
equivalent widths or the $3~\sigma$ equivalent width limits for the
displayed transitions.

\subsection{Galaxy Identification}
\label{id}
One potential candidate and one confirmed galaxy were found to be
associated with absorption systems seen toward PG~$0117+213$ by
\citet{csv96} (and references therein).  Both were imaged with
broadband filters ($g(4900/700)$, $R(6930/1500)$, $i(8000/1450)$, and
NICMOS) which allowed for the determination of the rest frame {\it B}
and {\it K} magnitudes.  The galaxy candidate thought to be associated
with the $z=0.5764$ absorption system has a {\it B} luminosity of
$2.32 L_B^{\star}$ (where $L_B^{\star}$ is the {\it B}-band Schechter
luminosity, and for $q_0=0.05$ and $h=H_0/100{\rm km s}^{-1}
\rm{Mpc}^{-1} = 1$ as it was tabulated in \citet{csv96}), a {\it B-K}
of $4.00$, and an impact parameter of $5.1~h^{-1}$~kpc.  This galaxy
candidate is too close to the quasar to be able to be classified
morphologically, though it has a very red color, meaning it could be
an elliptical galaxy or a very red spiral galaxy.  The confirmed
galaxy at $z=0.7290$ has a {\it B} luminosity of $3.27 L_B^{\star}$
(from \citet{csv96}, again using $q_0=0.05$ and $h=1$), a {\it B-K} of
$4.02$, and an impact parameter of $36.0~h^{-1}$~kpc.  The galaxy at
$z=0.7290$ is a face on, barred spiral; approximately SBa
\citep{glenn}.

There are three other galaxies within $10$~\arcsec of the quasar which
are potential candidates for the $z=1.0480$, $1.3250$, and $1.3430$
absorption systems, all $\sim3$ magnitudes fainter than the $3.27 L_B^{\star}$ galaxy
associated with the $z=0.7290$ absorber \citep{glenn}. In particular,
one galaxy candidate, if at $z=1.0480$, would have an impact parameter
of $22.7~h^{-1}$~kpc (using $q_0=0.05$ and $h=1$, as in \citet{csv96}, for
consistency), making it a likely candidate for the strong absorber.

\section{Modeling Techniques}
\label{mtech}
Through the use of photoionization and collisional ionization models,
we constrain the ionization conditions, metallicities, chemical
compositions, and kinematic makeup of five systems along the line of
sight toward PG~$0117+213$.  These methods are the same as were
employed in \citet{weak1634}, \citet{ding1634}, \citet{ding1206},
and \citet{zonak}. 

For all five systems, we began our modeling by focusing on the
absorption features in the Keck I/HIRES data, since they had the
highest signal-to-noise and resolution of the spectra available.  For
all but the primary component of the $z=0.5764$ system, we used the
{\MgIIdblt} transitions as a ``kinematic template''.  We found a
minimum number of Voigt profile components required to fit the
observed structure of the kinematic template transition.  For the
$z=0.5764$ system, we used the {\MgI}~$2853$ to determine a kinematic
template for fitting the primary component, since the {\MgII} profiles
were saturated.

For our photoionization models, we used an ionizing background
spectrum determined at the redshift of each system, based on the most
current Haardt-Madau extragalactic background radiation spectrum
available \citep{hm01}.  The spectrum for each redshift was determined
by linearly interpolating between the tabulated redshift values.  We
adopted a model ionizing spectrum that included contributions from
star-forming galaxies (with a relatively high escape fraction of
$10$\%) as well as from quasars.  The effect of absorption by the
intervening Ly$\alpha$ forest clouds was also included.  A quasar-only
spectrum would be harder and would generally lead to stronger high
ionization absorption.  We discuss the effects of using such a
spectrum in \S~\ref{spec}.

In general, it was not possible to produce the observed absorption in
all ionization stages, and at all velocities, from the original
kinematic template.  A minimum of one to three phases of gas were
required to fit all transitions in a given system.  Each phase can be
placed into one of four different regimes: low ionization, very-low
ionization, high ionization, or a general ``Other'' category.  Clouds
are defined as falling in the low ionization regime if the ionization
parameter, $U$, (defined as the ratio of the number density of
ionizing photons to the number density of hydrogen) is in the range
$-5 < \log U < -2.5$.  In this case, the majority of the {\MgII}
absorption would arise from low ionization clouds.  Clouds in the
very-low ionization regime, with $\log U < -5$, generally give rise to
significant amounts of absorption from neutral species.  High
ionization clouds are defined as those with $\log U > -2.5$; they may
have detected {\CIV}, {\NV}, and/or {\OVI} absorption, and may, in
principle be photoionized or collisionally ionized.  The ``Other''
category consists of clouds that do not fall cleanly into one of these
regimes, either because they fall in between regimes, because they are
collisionally ionized, or because they cannot be constrained well
enough to be classified.  The ionization fraction can be defined as
the ratio of neutral to total hydrogen and for our models this can be
read from Tables~\ref{val134}--\ref{val057}.  The ionization fraction
is closely related to $\log U$, but there can be a dependence on
metallicity.

\subsection{The Low Ionization Phase}
\label{modion}
Initially, we assumed that all of the observed {\MgI} was made in the
same phase as the {\MgII}.  We applied photoionization models to these
{\MgII} clouds, adjusting the parameters to fit the observed {\MgI},
as well as the other lower-ionization transitions, such as {\FeII},
{\OI}, {\SiII}, and {\CII}, when these transitions were covered.  For
most of our systems, this method provided adequate models of the
observed lower-ionization gas.  To constrain a model of a system, we
first used a minimum number of Voigt profiles to fit the {\MgII} (or
{\MgI} as mentioned above for the $z=0.5764$ system) \citep{cv01},
giving us a column density and Doppler parameter value for each
component.  We then used Cloudy, version $94.0$ \citep{cloudy}, with
various values of {\IP} and metallicity, to find expected column
densities for the other observed transitions.  The thermal Doppler
parameter for each chemical element was determined using the cloud
temperature given by Cloudy, and combined with the Doppler parameter
due to turbulence/bulk motion to obtain the total model Doppler
parameter for that element.  The turbulence/bulk motion Doppler
parameter was calculated based on the observed Doppler parameter for
the kinematic template transition. Synthetic line profiles based on
the model column densities and Doppler parameters were convolved with
the line-spread functions of the appropriate spectrograph, creating
models of the contribution of the low ionization phase to all observed
transitions.  We adjusted the {\IP} and metallicity to fit the
observed {\MgII}, {\MgI}, and {\FeII} profiles, and the available
{\Lya} and/or Lyman Limit features.  In many cases, these models
consistently produced the other observed low ionization transitions
(e.g. {\AlII}, {\CII}, and {\SiII}).

We found for our ``best-fit'' models metallicities in the range of
$-2.0{\le}\log{\frac{Z}{Z_\sun}}{\le}1.0$ for the low ionization
phase, while {\IPs} were in the range of $-5.0{\le}\log{U}{\le}-2.0$.
Our criterion for a ``best-fit'' model was that the model adequately
fit the largest number of transitions with the fewest number of
components.  In some cases, adjustments had to be made to the
abundance patterns for the model to correctly match all the observed
transitions.  Although we considered a strict $\chi^2$ statistic, we
found that individual pixels (which could be influenced by blends or
correlated noise) often dominate this value.  A ``by-eye'' comparison
of the models to the data generally produced {\IPs} and metallicities
accurate to $\sim0.1$~dex.  This process was illustrated in Fig.~6
of \citet{weak2}.  Our ``best-fit'' models of contributions
from the low ionization phase are superimposed (as dashed curves) on
the data in Figures \ref{fig134}-\ref{fig057}.

In one case (the $z=0.5764$ system), this method could not fit all of
the observed transitions, calling for the addition of a very-low
ionization phase.  In three of the systems ($z=1.3430$, $z=1.3250$,
and $z=0.5764$), higher ionization transitions, such as {\CIV}, {\NV},
and {\OVI}, were also observed, but could not be produced adequately
by the low ionization phase clouds (see the dot-dash curves in Figures
\ref{fig134}, \ref{fig132}, and \ref{fig057}).  This was sometimes due
to constraints on the {\IP} based on very-low and low ionization
transitions, and sometimes due to distinct differences between the
kinematic structure of the high and low ionization phase profiles.
This called for the addition of a high ionization phase in these three
cases.  In three systems ($z=1.0480$, $z=1.3250$ and $z=1.3430$), an
{\HI}-only cloud was needed to produce all the observed hydrogen
absorption.  In the first two cases, these {\HI} clouds were far
removed in velocity space from the rest of the absorption features.

\subsection{The Very-Low Ionization Phase}
\label{lowion}
For the primary absorption region of the $z=0.5764$ absorber, we were
unable to use {\MgII} as a kinematic template to determine {\IPs}
consistent with the other very-low and low ionization transitions (for
any assumed metallicity).  At first, we believed this problem to be a
result of ambiguity of the {\MgII} absorption strength, due to
saturation.  However, after exploring a wide range of Voigt profile
fits to the {\MgII}, we were unable to find any model that
satisfactorily fit the observed {\MgII}, {\MgI}, {\CaII} and {\TiII}.
We were able to reconcile this by adding a very-low ionization phase.
The kinematic template for this phase was drawn from a Voigt profile
fit to the {\MgI}.  Using this template we tried to determine {\IPs}
and metallicities that would produce the observed {\CaII}, {\TiII},
and {\FeII} absorption.  Over the full range of parameters, {\TiII}
and {\FeII} were underproduced, and abundance pattern variations were
considered.  The addition of a very-low ionization phase and the
optimal parameters given in the ``best fit'' model (Table
\ref{val057}) are consistent with the philosophy of minimizing the
required abundance pattern adjustments.  The contribution of the
very-low ionization phase can be seen as the dotted line in Figure
\ref{fig057}.  The low ionization phase for $z=0.5764$ was then
constrained by {\SiII}, {\MgII}, {\AlII}, {\AlIII}.

It is possible that very-low ionization phases would exist in other
systems, however they are not necessary to explain absorption in the
observed transitions in those cases.  If a very-low ionization phase
did exist in one of the other systems, the parameters of its low
ionization phase would have to be adjusted to compensate.

\subsection{The High Ionization Phase}
\label{highion}
After the very-low and low ionization transitions had been modeled,
the {\OVI}, {\NV}, and {\CIV} absorption of three systems ($z=1.3430$,
$z=1.3250$, and $z=0.5764$) was still underproduced.  In the case of
the $z=0.5764$ system, the higher ionization absorption at the same
velocity as the low ionization absorption (at $v=-30$ and $-9$~{\kms})
could be produced by the low ionization phase.  However, for the high
ionization absorption observed at other velocities, as well as the
absorption in the other two systems, the kinematic profiles of the
high ionization absorption were so different from that of the
underlying low ionization absorption that they could only be fit by
adding a high ionization phase.  In the case of the $z=0.5764$ system,
the high ionization phase was unconstrained, apart from the lack of
production of low ionization transitions.  For the $z=1.3250$ and
$z=1.3430$ systems, this phase was constrained by {\CIV}, {\NV}, and
{\OVI}.

In principle, either photoionization or collisional ionization could
determine the ionization balance for the high ionization phase of a
given component.  We were able to differentiate between systems that
required collisional ionization models and those that could be
consistent with photoionization models by comparing the different
absorption patterns each creates.  Collisionally ionized clouds tend
to produce a narrow range of ionized transitions (e.g. {\CIV} or {\NV}
or {\OVI}, but never all three), whereas photoionized clouds tend to
have broader ranges.  Our high ionization components are shown in the
figures as the dot-dashed line.
The significance of the lack of a high ionization phase in the
$z=1.0480$ and $z=0.7290$ systems will be discussed in \S~\ref{dis104}
and \S~\ref{dis072}.

An illustration of how the parameters of high ionization components
are typically constrained by our modeling technique was given in
\citet{zonak}.  Since the high ionization gas is in the optically
thin regime, the ionization parameter can be determined from the
ratio of any two high ionization transitions, independent of
metallicity.  The metallicity is then constrained, for that ionization
parameter, based upon the {\Lya} and {\Lyb} transitions, particularly
for our $z=1.3430$ and $z=1.3250$ systems.

\subsection{Other Phases}
\label{otherion}
Two of our systems, $z=1.0480$ and $z=1.3250$, required an additional
{\HI}-only cloud to be added to fill out the observed {\Lya} profile.
In both cases, the cloud was added to fill out the blueward wing of
the {\Lya} profile, at $v=-270~{\kms}$ for the $z=1.0480$ system, and
at $v=-140~{\kms}$ for the $z=1.3250$ system.  In order to get a rough
idea on how the excess {\Lya} absorption might be produced, we
considered a simple single {\HI}-only cloud model for each absorbers.
The metallicities of these clouds were required to be low in order not
to produce metal lines at the same velocities.

The $z=1.3430$ system also required an additional {\Lya} component,
however this cloud was a broad absorber, covering the full velocity
range of the system.  It primarily produced the wings of the {\Lya}
profile.

A {\SiIII} component was added to the $z=1.0480$ system to fill out the
{\SiIII}~$1207$ and {\Lya} absorption at $v\sim-200~{\kms}$.  This helped
produce much of the {\Lya} absorption in that velocity range, but a
{\HI}-only cloud was still required for this system, as discussed above.

Also, a collisionally ionized phase was added into the $z=1.0480$
system in order to fit the excess {\SiIV} absorption observed at a
position of $6$~{\kms}.  Although the {\SiIV}~$1394$ is blended from
$-300~{\le}~v\le100$~{\kms}, the {\SiIV}~$1403$ matches the main spike
of the {\SiIV}~$1394$ at $v=6$~{\kms}, giving us reason to believe
that at least that small part is a real feature.  Our reasons for
choosing a collisionally ionized phase, as opposed to a photoionized
phase, will be discussed in \S\ref{res104}.

\subsection{Input Spectra Variations}
\label{spec}

We used an input spectrum which is an updated version of the spectrum
presented in \citet{hm}.  We obtained the new spectrum from
\citet{hm01}, which was an extragalactic background radiation spectrum
from quasars, similar to the original, but with additional
contributions from star forming galaxies with an escape fraction of
$10$\%.  We chose to use this spectrum, as opposed to the older one,
because our systems exist during an epoch of increased star formation
($z\approx1$) so that we would expect some stellar contribution.  The
primary effect of using a QSO-only spectrum is an increase in high
ionization absorption contributed by the lower ionization clouds,
which is a result of this spectrum being harder than the spectrum with
star forming galaxies.  As we discuss in \S~\ref{results}, all of our
models either produced too much high ionization absorption to begin
with, or had the high ionization gas so far offset from the lower
ionization gas that making more in the lower ionization phases would
not adequately model the data.  A more thorough study of the effects
of varying the spectral shape on models of some other absorption
systems is presented in \citet{ding04}.

\section{Model Results and Discussion}
\label{results}

We now present our numerical constraints on the properties of the five
{\MgII} systems we analyzed in the line of sight toward PG~$0117+213$.
Photoionization modeling was used to determine the range of acceptable
ionization parameters ($\log U$) and metallicites ($\log Z$), when
adequate constraints were available.  Collisional ionization was also
considered and temperatures ($\log T$) were constrained for possible
collisional models.  The transitions used to constrain these values
are listed, when appropriate.  The sample models were overlaid on the
data, and are presented in Figures \ref{fig134}-\ref{fig057}.  The
equivalent widths of key transitions of each system are given in
Tables \ref{ew134}-\ref{ew057}.  The input parameters used to create
our models, as well as some of the derived parameters are given in
Tables \ref{val134}-\ref{val057}.  The tick marks in the figures
indicate the locations of model clouds from the tables, as noted
specifically in the figure captions.
The redshifts given for each system
are optical depth weighted mean values derived from the {\MgII}
transitions observed by Keck I/HIRES.  Because the Lyman limit break
due to the $z=1.3430$ system affects the modeling of the $z=1.3250$
system, we present results starting with the highest redshift system
and ending with the lowest.

\subsection{The $z=1.3430$ System}
\subsubsection{Results}
\label{res134}

\begin{figure*}
\figurenum{1}
\epsscale{0.8}
\plotone{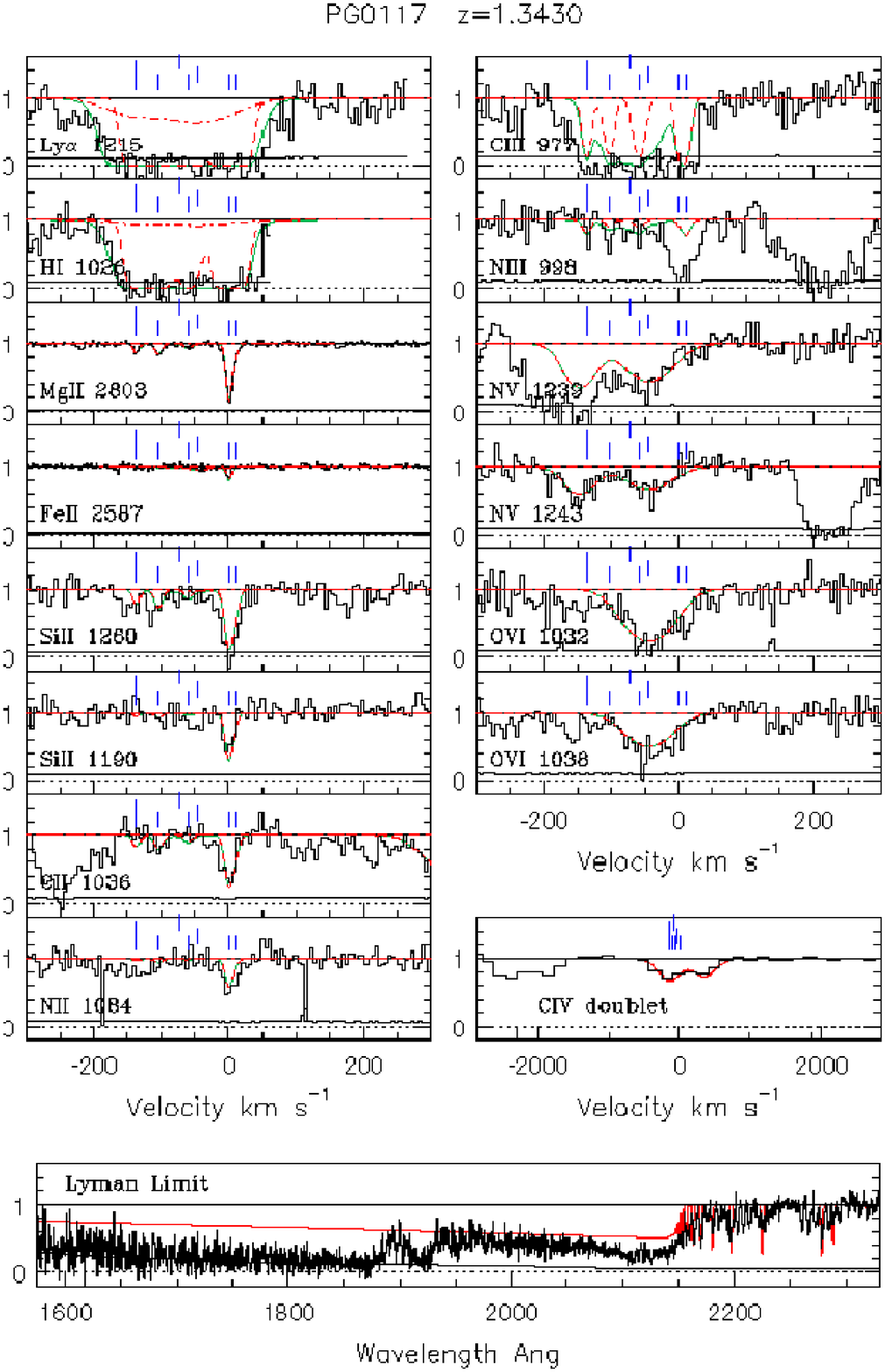}
\protect\caption{
\baselineskip = 0.7\baselineskip
\scriptsize{
The transitions observed for the $z=1.3430$ system are presented here
in velocity space, centered around the {\MgII} optical depth
center, as determined in \citet{cv01}.  All transitions were observed
with {\it HST}/STIS, except {\CIV}, which was covered in a Palomar
$200~$inch spectrum \citep{ss92}. The bottom row of ticks mark
positions of model clouds {\MgII}$_1$--{\MgII}$_5$, the middle row
represents clouds {\OVI}$_1$ and {\CIV}$_1$, and the top row
marks cloud {\Lya}$_1$ from Table~\ref{val134}. The histogram shows the
data, with the black line near zero showing the $1~\sigma$ error.  The
smooth curve indicates the full model as described in
\S~\ref{res134}.  Ionization phase contributions are shown (where
applicable) with the dotted line (very-low), the dashed line
(low), and the dot-dashed line (high).  The lower panel presents the
region of the {\it HST}/FOS spectrum covering the partial Lyman
limit break (at $\sim 2150$~{\AA}, with the model contribution from
this system only superimposed (the $z=1.3250$ system also contributes
to the break).  } }
\label{fig134}
\end{figure*}

Figure \ref{fig134} shows key transitions covered in the Keck I/HIRES and
{\it HST}/STIS spectra for this system.  The {\MgII}~2803 in the Keck I/HIRES
spectrum was fit with a total of five Voigt profile components from
$-137$~{\kms} to $11$~{\kms} \citep{cvc03}, clouds {\MgII}$_1$--{\MgII}$_5$,
given in Table~\ref{val134}.  The rest frame equivalent width of {\MgII}~2803
is $W_r(2803) = 0.147\pm0.004$~\AA~\citep{weak1}.  {\MgII}~2796 fell off an
echelle order, so we do not know its equivalent width, but it is likely that
this is a multiple cloud, weak {\MgII} absorber, below the
$W_r(2796)=0.3$~\AA~ cutoff for strong systems.  The strongest component was
blended with a weaker component to the red, giving rise to an asymmetric
profile.  {\FeII} is detected only in the strongest component (cloud
{\MgII}$_4$ at $v=1$~{\kms}).  In the STIS spectrum, {\SiII}~1260 and
{\SiII}~1190 are detected, but the former is too strong relative to the
latter to be consistent with a Voigt profile fit.  {\SiII}~1193 cannot be
used as a constraint because it is blended with Galactic {\MgII}~2796.  We
adopt {\SiII}~1190 as the most adequate constraint on $N({\SiII})$.
{\CII}~1036 is detected in the STIS spectrum.  Unfortunately, the stronger
transition, {\CII}~1335, is not covered.  {\NII}~1084 is affected by a blend
with {\OVI}~1038 at $z=1.4477$.

The intermediate ionization transitions for this system are strongly
affected by blends.  {\SiIII}~1207 is blended with the {\Lya} line
from the $z=1.3250$ system, and cannot be studied.  A strong line is
detected at the expected position of {\CIII}~977.  This line seems
quite strong relative to the other lines from this system, and could
suffer from a blend.  However, we could not confirm that the blend is
with a {\Lya} line at $z=0.8826$ because {\CIV}~1551 for that system
is not detected and {\Lyb} is not covered in the spectrum.  We will
consider both the possibility that all of the absorption is
{\CIII}~977 at $z=1.3430$ and the possibility that there is a blend
affecting it.  {\NIII}~989 may also be blended with a {\Lya} line at
$z=0.9077$, but there is nothing to confirm this identification.

The high ionization transitions, {\NVdblt} and {\OVIdblt} are detected in the
STIS spectrum. Equivalent widths for these transitions are listed in
Table~\ref{ew134}.  The {\NV}~1239 transition is blended with {\Lya} from a
system at $z=1.3866$, but the {\NV}~1243 transition appears to be relatively
unaffected by blends.  {\NV} is detected in components centered at
$v=-147$~{\kms} and $v=-44$~{\kms}, clouds {\OVI}$_1$ and {\CIV}$_1$ in
Table~\ref{val134}.  {\OVI}~$1032$ is also detected in cloud {\CIV}$_1$ at
$v=-147$~{\kms}, but it is unclear if the detection blueward of $-100$~{\kms}
is real or a blend.  A blend with {\Lyg} from a $z=1.4986$ system is known to
affect {\OVI}~1038 in that region.  {\CIVdblt} is detected in a low
resolution FOS spectrum.  The {\CIV} is strongest at wavelengths
corresponding to the blueward component in {\NV}, and not to the strongest
low ionization component.

The metallicity of the low ionization phase of this system can be
constrained by {\Lya} and {\Lyb} as well as by a partial Lyman limit
break observed at the corresponding wavelength in the lower resolution
FOS spectrum, also shown in Figure~\ref{fig134}.  The situation is
significantly complicated by the fact that the breaks due to the
$z=1.3250$ and $z=1.3430$ systems are superimposed.  To make matters
worse, the continuum fit in the region above the partial Lyman limit
break is highly uncertain, with different fits giving an optical depth
range of $\tau\approx1.3\pm0.2$, to one sigma.  We derive constraints
on the $z=1.3430$ system, taking into account the uncertainty in the
continuum fit, and assuming that none or all of the break could arise
from it.  Then we apply these constraints to find a consistent
solution for the $z=1.3250$ system.

We begin with the five component Voigt profile fit to {\MgII}~2803.
{\FeII}~$2383$ provides a constraint on the {\IP} of $-3.9$ to $-4.0$
for the main component, cloud {\MgII}$_4$.  The other four low
ionization clouds do not have well constrained {\IPs}, with only weak
upper limits provided by {\NII}~$989$ and {\CII}~$1335$.  We pushed
the {\IP} values for these clouds ({\MgII}$_1$--{\MgII}$_3$ and
{\MgII}$_5$) as high as possible in order to maximize the production
of {\CIII}~$977$.  The contribution of this phase of gas is shown as a
dashed line on Figure~\ref{fig134}.  High ionization transitions are
not significantly produced in this phase.

Next we consider metallicity constraints for the {\MgII} clouds.  We
simplified our models by assuming that all the {\MgII} clouds in this system
({\MgII}$_1$--{\MgII}$_5$) had the same metallicity.  The simplest model of
the Lyman limit break would have the $z=1.3430$ system produce the full Lyman
limit break absorption.  This model would also produce most of the {\Lyb}
absorption in the same clouds.  For the {\Lyb}, it is possible to fit the
width of the saturated trough with the narrow, low ionization clouds, but it
is not possible to fit the blue wing.  {\Lya} still requires additional broad
components to fit both wings.  All five clouds have a metallicity of
$\sim-0.8$ in a model that best fits {\Lyb} and the partial Lyman limit
break, but the uncertainty is large ($\pm 0.2$ or $0.3$~dex) due to the large
uncertainty in the continuum fit.  However, if we let the $z=1.3250$ system
have a contribution to the Lyman limit break that is comparable to the
$z=1.3430$ system, we find a metallicity value of $\sim-0.3$ for both
systems, with the same uncertainty of $\pm 0.2$ or $0.3$~dex.  An additional
{\Lya} component is still required for the $z=1.3430$ system in this case,
but the model fits the Lyman limit break more consistently in the latter
case, and so we use this case for the model displayed in Figure~\ref{fig134},
as well as the numbers given in Table~\ref{val134}.  This finding supports
the predictions of \citet{archiveI}, who found that the metallicities of
these two systems should be equal, through modeling of a lower resolution
spectrum of the Lyman limit break.

The contributions of this $\log Z = -0.3$ model, for the low ionization
clouds, to {\Lya}, {\Lyb}, and the partial Lyman limit break are shown in
Figure \ref{fig134}.  Since we have assumed the same metallicity for the five
low ionization clouds, among the five, the main component at $1$~{\kms} is the
dominant contributor to the Lyman limit break.

In order to fill in the {\Lya} profile, an additional broad {\Lya}
component, {\Lya}$_1$ in Table~\ref{val134}, is required.  This cloud
was placed roughly at the center of the main {\Lya} absorption
($z\simeq-73~${\kms}), and is constrained to have low metallicity
($\log Z < -2.5$) so as not to produce metal transitions.  In order to
correctly produce the observed absorption wings, a minimum Doppler
parameter of $b\approx45$~{\kms} is required, which leads to a minimum
column density of $\log~N=16.6$~{\cmsq} (we adopted
$\log~N=16.8$~{\cmsq} in Table~\ref{val134} and Figure~\ref{fig134},
since this model provides a slightly better fit to the {\Lya}
profile).  This column density is larger that $N({\HI})$ for the low
ionization clouds, so that this broad {\Lya} component dominates the
Lyman limit break for this system.

The high ionization transitions ({\CIV}, {\NV}, and {\OVI}) cannot be
produced by the {\MgII} clouds.  They do not occur at the same
velocities as the strong, low ionization absorption.  The redward high
ionization component at $v=-44$~{\kms} (cloud {\OVI}$_1$) has detected
{\NV} and {\OVI}.  Photoionization models with any {\IP} would
overproduce {\CIV} at that velocity.  A collisional ionization model
with $\log T\sim5.33$ provides the best fit to the {\NV} and {\OVI} at
$v=-44$~{\kms}.  For the blueward component at $v=-147$~{\kms} (cloud
{\CIV}$_1$) {\CIV} and {\NV} are detected, and {\OVI} is ambiguous.  A
photoionization model would overproduce {\CIII} blueward of
$-150$~{\kms}, beyond the saturated trough that may or may not be
{\CIII}~977 from this system.  A collisional ionization model with
$\log T\sim5.12$ can fit the observed {\CIV} and {\NV} absorption
without overproducing other transitions.

It is possible to tune the metallicities of the two broad collisional
components so that the wings of the {\Lya} and {\Lyb} lines are fit.
For the $\log T\sim5.33$ cloud at $v=-44$~{\kms} (cloud {\OVI}$_1$
that produces {\OVI} and {\NV} absorption), this requires $\log Z \sim
-0.2$.  For the $\log T\sim5.12$ cloud at $v=-147$~{\kms} (cloud
{\CIV}$_1$ that produces {\NV} and {\CIV} absorption), this requires a
supersolar metallicity of $\log Z
\sim 1.0$.  These clouds are shown in the final, full profile plotted
in Figure \ref{fig134}, but not as individual lines.

\subsubsection{Discussion}
\label{dis134}

The $z=1.3430$ system has low ionization absorption clouds
({\MgII}$_1$--{\MgII}$_5$) that appear to have densities
($\log~n_H\sim-2$) and temperatures ($T\sim10000$~K) consistent
with warm ISM material (\cite{bregman}, and references therein).
However, the high ionization absorption is not aligned with the lower
ionization absorption, indicating that it is not caused by the same
clouds.  Blending is a large problem with the high ionization
transitions, but there is enough overlapping unblended data between
the {\CIV} and {\OVI} doublets to allow for confirmation of our
models.  These high ionization clouds favor collisional ionization at
$T\sim100000-200000~$K, which could be caused by shock heating of the
region.  This could be the same mechanism as the conductive interface
model proposed by \citet{fox}.  The different high ionization
components show different metallicities, with the red component (cloud
{\OVI}$_1$) having roughly the same metallicity as the lower
ionization transitions (${\log}Z\sim-0.2$).  The blue component (cloud
{\CIV}$_1$), however, has a supersolar metallicity.  This is unusual
and perhaps due to ambiguities in the model, but it could also
indicate a region of intense enrichment by supernovae.

This system, while just below the cutoff to be classified as a strong
{\MgII} absorber, has kinematics very similar to classic strong
systems \citep{kinmod, cv01}.  It also shows a partial Lyman limit
break, in contrast to the classic strong system's full Lyman limit
break.  This may indicate a scenario in which this line of sight
skirts the edge of, or a sparse region of, a galaxy disk
that would otherwise create a strong {\MgII} absorption system.

Having high ionization gas in a phase that is different from the lower
ionization gas is not unusual for strong or weak {\MgII} systems
\citep{weak1,archiveII,weak2,weak1634}.  When found in a strong system, the
high ionization absorption usually covers similar velocities as the
lower ionization absorption.  It is either kinematically grouped with
the low ionization gas or in a broad profile surrounding the low
ionization component \citep{ding04,ding1206}.  However, in this
multiple cloud weak {\MgII} system at $z=1.3430$, the high ionization
clouds are kinematically quite distinct from the low ionization
clouds.  There are other cases of multiple cloud weak systems showing
such distinct high ionization features \citep{zonak,ding04}.  If this
trend is still apparent in a larger sample, it could be very telling
about the differences between multiple cloud strong and multiple cloud
weak {\MgII} systems.  One example is the $z=1.04$ system toward
quasar PG~$1634+706$ \citep{zonak}.  The authors of that paper found
that their system had kinematically offset {\OVI} components
$\sim50$~{\kms} to the red and $\sim50$~{\kms} to the blue of the main
{\MgII} clouds.  The authors suggested that their system could be
related to dwarf galaxy winds.  The broad {\OVI} clouds also had a
significantly higher metallicity than the {\MgII} clouds.  Our
$z=1.3430$ system could be produced by a similar situation, though its
{\OVI} appears to be collisionally ionized, not photoionized.  It is
possible for winds to have a higher metallicity than that of the host
galaxy \citep{martin}.  These low ionization clouds could be
supernova-enhanced host galaxy material, material entrained in the
outflowing winds, or material in a conductive interface zone.

\subsection{The $z=1.3250$ System}
\subsubsection{Results}
\label{res132}

\begin{figure*}
\figurenum{2}
\epsscale{0.8}
\plotone{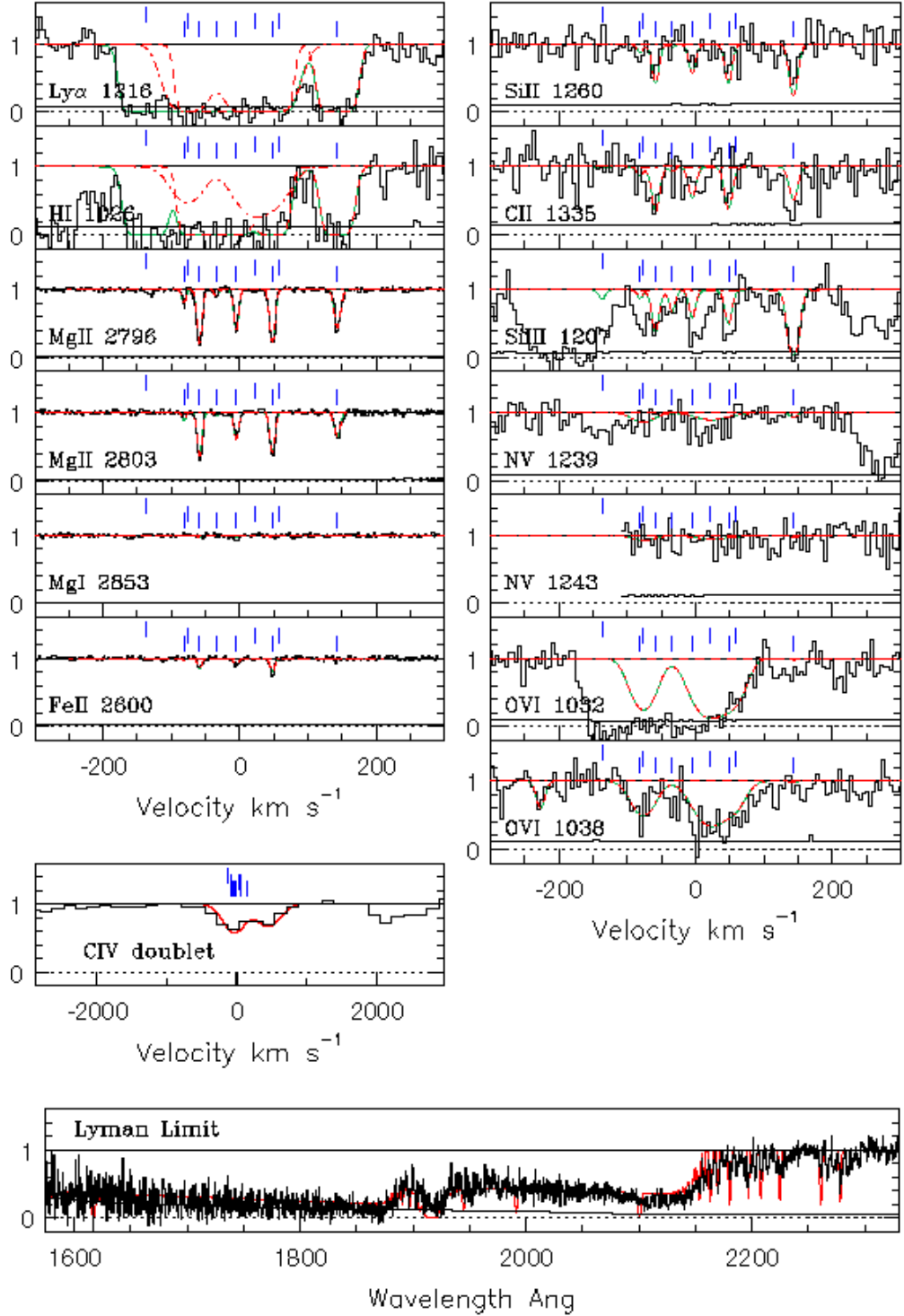}
\protect\caption{
\baselineskip = 0.7\baselineskip
The same as Fig~\ref{fig134}, except for the $z=1.3250$ system.
The lower row of ticks marks the low ionization phase clouds,
{\MgII}$_1$--{\MgII}$_5$, the middle row represents the high
ionization phase, {\OVI}$_1$--{\OVI}$_3$, and the top row
marks the additional cloud, {\Lya}$_1$ (see Table~\ref{val132}).
The lower panel is again the {\it HST}/FOS spectrum, but now
the combined model contributions of both this system and the $z=1.3430$
system are superimposed.
}
\label{fig132}
\end{figure*}

The {\MgIIdblt} for this system, shown in Figure \ref{fig132} can be
fit with six distinct components from $-82$~{\kms} to $143$~{\kms},
clouds {\MgII}$_1$--{\MgII}$_5$ in Table~\ref{val132} \citep{cvc03}.
The total equivalent width of the system of $W(2796) =
0.291\pm0.011$~\AA~ classifies it as a multiple-cloud, weak {\MgII}
absorber, just below the $0.3$~\AA~ threshold for strong systems.
{\MgI} and {\FeII} were also detected in the HIRES spectrum, while
{\Lya}, {\Lyb}, {\SiII}, {\CII}, {\SiIII}, and {\OVI} were detected in
the STIS spectrum.  There is Galactic {\MgII}~2803 $\sim200$~{\kms}
blueward of the {\SiIII}~1207 detection, and an unidentified feature
$\sim250$~{\kms} redward of it.  It it likely that there are some
unidentified absorbers blending with {\SiIII}~1207 in the region of
interest as well.  It is clear that {\NV}~1239 is at least partially
blended, since {\NV}~1243 is not detected at $v < -180$~{\kms}.
{\OVI}~1032 is blended with {\Lyb} from an {\OVI} system at
$z=1.3385$, but the region at $v>40$~{\kms} ($v>200$~{\kms} at
$z=1.3385$) is not affected by this blend since the corresponding
{\Lya} is not detected.  {\OVI}~1038 at $v>0$~{\kms} is blended with
the {\OVI}~1032 line from the $z=1.3385$ system.  Taking into account
these blends, the {\OVI} doublet can be fit with a minimum of three
Voigt profile components, shown in Figure~\ref{fig132} and listed
as clouds {\OVI}$_1$--{\OVI}$_3$ in Table~\ref{val132}.

We first consider whether the detected {\CIV} and {\OVI} absorption
can arise in the same phase of gas as the {\MgII} and other lower
ionization transitions.  The ionization parameters of the three
{\MgII} clouds with detected {\FeII} (clouds {\MgII}$_2$, {\MgII}$_4$,
and {\MgII}$_5$) are all in the range $-3.6 < \log
U < -3.3$, while upper limits can be set for the other three.  With
these ionization parameters, tuned to fit {\FeII}, {\SiII} is
overproduced.  Upper limits for the ionization parameters of the
three clouds without detected {\FeII} (clouds {\MgII}$_1$, {\MgII}$_3$,
and {\MgII}$_6$) are derived such that {\CII},
{\SiII}, and {\SiIII} are not overproduced.  A model using these
limits, given in Table \ref{val132}, severely underproduces the
{\CIV} and {\OVI} absorption, as shown by the dashed line in Figure
\ref{fig132}.

The metallicity of the {\MgII} clouds ({\MgII}$_1$--{\MgII}$_6$) was
constrained by the partial Lyman Limit break and the {\Lya} and {\Lyb}
profiles.  Since the Lyman limit breaks of the $z=1.3250$ and
$z=1.3430$ systems overlap, an iterative process was needed to
co-model these two systems.  For this $z=1.3250$ system, the {\Lya}
and {\Lyb} profiles are best fit if clouds {\MgII}$_5$ at
$v=48$~{\kms} and {\MgII}$_6$ at $v=143$~{\kms} have metallicities of
$\log Z \sim-0.3$.  Smaller metallicities for these two clouds are
ruled out, since {\Lya} and {\Lyb} would be overproduced.  The
metallicities of clouds {\MgII}$_1$--{\MgII}$_4$, at $v=-82$, $-60$,
$-35$, and $-5$~{\kms}, are not constrained in this way, and so we assume
that they also have metallicities of $\log Z\sim-0.3$.  Even for
extremely low metallicities we could not fit the blue edge of the
{\Lya} and {\Lyb} profiles, and for such low metallicities the Lyman
limit break constraint would be severely violated.  Combined with the
{\Lya} component (cloud {\Lya}$_1$ discussed below), these {\MgII} clouds fill out the
remainder of the Lyman limit break left after the $z=1.3430$ system
was modeled.  The clouds {\MgII}$_2$ and {\MgII}$_5$, at $v=-60$ and $48$~{\kms},
were the primary contributors to the Lyman limit break, among the metal lines.

Also detected was {\MgI}~2853 at a $\sim 3\sigma$ level from the
$v=-5$~{\kms} cloud, {\MgII}$_4$.  It is not fit by the low ionization clouds used
to fit {\MgII}, {\FeII}, {\SiII}, and {\CII}.  This may require a
separate very-low ionization phase \citep{ding1634}.  Properties of
such a phase cannot be constrained by the present data.

The {\CIV} and {\OVI} can both be produced with three additional
clouds, {\OVI}$_1$--{\OVI}$_3$, which were optimized on {\OVI} based on the {\OVI} Voigt
profile fit described above (and shown by the dot-dashed line in
Figure~\ref{fig132}).  The best fit to the low resolution {\CIV}
profiles is obtained for ionization parameters of $\log U\sim-1.0$ for
clouds {\OVI}$_1$--{\OVI}$_3$.  The metallicities of all the {\OVI} clouds were
assumed to be the same, yielding a value of $\log Z\sim-0.3$, based on the
first red wing of the {\Lyb} profile.  This leads to cloud sizes of
$5$--$20$~kpc.  Nitrogen had to be lowered $0.8$ dex, $0.5$ dex, and
$1.2$ dex respectively from solar in our models of these three high
ionization clouds to prevent overproduction of {\NV}, which has no
absorption observed.  These clouds account for much of the {\Lya} and
{\Lyb} absorption, as shown in Figure \ref{fig132}, however they do
not fully account for the {\Lya} and {\Lyb} at
$-180\le~$v$~\le-100$~{\kms}.

A final cloud centered at $-137$~{\kms}, {\Lya}$_1$, was added to fill
in the section of the {\Lya} profile that the {\OVI} could not account
for, as well as much of the {\Lyb} feature.  The metallicity and
ionization parameter of this cloud were only constrained so that there
was no production of any metal-line absorption.  A value of $\log
Z=-3.5$ was used in the sample model in Table \ref{val132}.  Due to
the lack of any constraints, cloud {\Lya}$_1$ was assumed to have an
{\IP} of $\log U=-2.0$.  For this model, the {\Lyb} profile is not
completely filled in if the {\Lya} profile is matched, especially on
the blue wing.  Either, the {\Lyb} is blended with unidentified lines,
or a more complex model is needed.  The {\Lya} also has a few sections
($v\sim100$~{\kms}, $v\sim190$~{\kms}) which are underproduced by the
model, but they could not be fit without overproducing {\Lyb}.  This
{\Lya} cloud ({\Lya}$_1$), along with the six {\MgII} clouds in this
system ({\MgII}$_1$--{\MgII}$_6$), produces the remaining Lyman limit
break absorption not filled in by the $z=1.3430$ system.

\subsubsection{Discussion}
\label{dis132}

It is difficult to distinguish disk material from halo material.
Inferences could be drawn from a statistically large sample, but for
any one system there is no way to make a clear separation because halo
clouds overlap kinematically with disk clouds.  However based on the
results of \citet{kinmod} and \citet{steidel02}, we statistically can say
that because of the lack of primary component, there is a low
probability that this line of sight passes through any significant
galaxy disk material.

The high ionization transitions in this system, in particular the
{\OVI}, are corrupted by strong blends, which makes analysis much more
difficult.  Still, we are able to confidently say that any high
ionization gas present must be independent of the low ionization gas
observed.  This is very similar to what we observed in the $z=1.3430$
system, but again in contrast with the standard picture of strong
{\MgII} systems.  Are these two systems, along with the $z=1.04$
system toward PG~$1634+706$ \citep{zonak}, prototype members of a
separate class of {\MgII} absorber?  This would be a class of multiple
cloud weak absorbers with unrelated high ionization kinematics.

The {\MgII} kinematics of the $z=1.3250$ system is very similar to a
subsystem of the $z=0.9254$ absorber presented by \citet{ding1206}
(called ``System A'' in that paper).  Alone, System A would be
classified as a multiple cloud weak {\MgII} absorber, but it is
clustered with Systems B ($z=0.9276$) and C ($z=0.9342$), a few $100$
and $1000$~{\kms} to the red, respectively.  Therefore, it is part of
what has been classified as a ``double'' strong {\MgII} absorber.
Though the metallicity of System A ($\log Z=0.5$) is higher than our
$z=1.3250$ system ($\log Z=-0.3$), the kinematics of the {\MgII} is
very similar, suggesting a similar environment for both.  The
$z=0.9254$ system is likely to be associated with either a $2~L_\star$
spiral galaxy, or a $0.2~L_\star$ galaxy with unknown morphology.
Both were found to be at the same redshift as the absorber, and at
about equal impact parameters ($43~h^{-1}$~kpc).  System B is
identified with an $L_\star$ galaxy at an impact parameter of
$38~h^{-1}~$kpc.  The whole situation is suggestive of a group of
galaxies with lines of sight at large impact parameter through the
outskirts of the galaxies.  This could be a hint of the galaxy type
and environment of the $z=1.3250$ system, for which we have no direct
information.

However, not all is the same between System A and the $z=1.3250$
absorber.  System A has very strong {\NV} absorption, whereas our
system has so little that an abundance pattern adjustment is required
to account for its deficiency.  While there have been many papers
which discuss strong nitrogen underabundances in DLAs, many of those
focus on systems with metallicites lower than our system ($-2<\log
Z<-1$ compared to our $\log Z=-0.3$ \citep{pettini}). Our system is
apparently different from such DLAs and from dwarf galaxies, which are
also nitrogen deficient, but have low metallicity as well
\citep{mateo}.  This may suggest an enrichment scenario which is only
primary, however coverage of oxygen transitions would be required to
be able to determine this.

A more interesting (though speculative) hypothesis is that our
$z=1.3250$ system and System A of \citet{ding1206} represent the
absorption features of the tidal tails and debris from interacting
galaxies.  Due to multiple interactions in a group, material may be
dispersed, with some regions collapsing to form star clusters, tidal
dwarf galaxies and other gaseous concentrations.  Should our line of
sight pass through such a region, pockets of higher density would be
randomly crossed, giving absorption profiles with many smaller
features, but no main feature.  Such a circumstance was suggested by
the group environment known for the PG~$1206+459$ absorbers
\citep{ding1206}, but in our case, there is much less supporting
evidence.  Unless the line of sight fell along one of the tails,
multiple interactions would be needed to produce the structures with a
large spread in velocity.  Without an image deep enough to resolve the
galaxy/galaxies associated with this absorber, we cannot test this
speculation.

There is also a small amount of {\MgI} absorption observed for the
$v=-5~${\kms} cloud, which our models cannot account for.  This
absorption cannot be produced in the same phase as the {\MgII}, and
can only be explained by an additional, very-low ionization phase at
the same velocity.  One way to create a phase like this is to have a
cloud with a very small Doppler parameter, placing the {\MgII} on the
flat part of the curve of growth, while {\MgI} is on the linear part
\citep{ding1634}.  Thus, the {\MgII} associated with the {\MgI} does
not have to be the strongest observed, as in this case.  This would
imply that the {\MgI} forms in smaller, dense regions of the {\MgII}
cloud, so that a single line of sight may only pass through one such
region.

\subsection{The $z=1.0480$ System}
\subsubsection{Results}
\label{res104}

The {\MgIIdblt} profiles for the $z=1.0480$ system are fit with seven
Voigt profile components, clouds {\MgII}$_1$--{\MgII}$_7$, in two
separate subsystems at $\sim 0$~{\kms} and $\sim -90$~{\kms}
\citep{cvc03}.  {\MgI}~2853 and {\FeII}~2600 are also detected in the
HIRES spectrum.  In the STIS spectrum, the {\Lya} profile is
significantly extended blueward of the two subsystems.  The {\SiII}
and {\CII} profiles have roughly the same shapes as the {\MgIIdblt},
as does the {\SiIII}, but the latter is contaminated by an unknown
blend (probably {\Lya} at $z\approx1.0325$).  We will consider the
possibility that the {\SiIII}~1207 at $v = -186$~{\kms} is a detection
rather than a blend, and that it produces {\Lya} centered at that
velocity.  The {\SiIV}~1394 is blended with Galactic {\MgI}~2853, but
the {\SiIV}~1403 is detected from the subsystem at $\sim 0$~{\kms}.
The {\NVdblt} was also covered in the STIS spectrum, but it was not
detected.  Finally, {\CIV} was covered, but not detected in a low
resolution optical spectrum from Palomar \citep{ss92}.

The {\IP} was first constrained for the seven {\MgII} clouds, assuming
that {\MgII}, {\FeII}, and {\MgI} are present in the same phase.  The
ionization parameters found for these clouds, {\MgII}$_1$--{\MgII}$_7$
in Table~\ref{val104}, ranged from $\log U = -4.1$ to $\log U = -3.6$
(see Table \ref{val104}).  This conclusion is nearly independent of
metallicity constraints, which are discussed below.  For clouds
{\MgII}$_1$, {\MgII}$_3$, and {\MgII}$_7$, those for which {\FeII} is
not detected, these $\log U$ values can be taken as lower limits.  For
these values of $\log U$, the {\SiIV} is not fully produced.  Since the
determined value was a lower limit, the ionization parameter of
cloud {\MgII}$_3$, at $v=-23$~{\kms}, could be raised to $\log U = -2.5$ so that it
does produce the {\SiIV} at that velocity.  However, an additional
phase would still be required to produce the {\SiIV}~1403 centered at
$v\sim6$~{\kms} (recall that the {\SiIV}~1394 is contaminated by a
blend at this velocity).

We consider photoionization and collisional ionization for the phase
of gas that produces the {\SiIV} absorption at $v\sim6$~{\kms}.  If it
is photoionized, either the {\SiIII} or the {\CIV} would be
drastically overproduced.  Such a case, for which a model needs to
dominantly produce a particular ionization stage, suggests collisional
ionization at a higher temperature.  Using {\SiIV}, {\SiIII}, {\CII},
and {\SiII}, the temperature was found to be $\log T\sim4.74$~K.
The parameters for this collisionally ionized cloud, {\SiIV}$_1$, are
listed in Table~\ref{val104}.

The possible {\SiIII} absorption at $v=-186$~{\kms} is centered on a
region of the {\Lya} profile that could be interpreted as a separate
component (from $\sim-250$ to $-130$~{\kms}).  It is possible to fit
this region of the {\Lya} profile and the {\SiIII} absorption at
$v=-186$~{\kms} with a single cloud, {\SiIII}$_1$, with $\log U\sim-2.1$
and $\log Z\sim-1.4$.  For this
metallicity, values of $\log U$ much smaller than $-2.1$ are not
possible because lower ionization transitions are not detected at this
velocity.  This model component was considered because of the small
amounts of flux apparently detected in the {\Lya} profile at
$v\sim-250$ and $v\sim-140$~{\kms}.  If this is valid, another {\Lya}
component, {\Lya}$_1$, without corresponding detected metal line transitions,
would have to be added at $v\sim-266$~{\kms}.  This component,
assuming the same metallicity as the {\SiIII} component, could not
have an {\IP} larger than $\log U\sim-3.0$ without overproducing
{\SiIII} absorption.  If the possible {\SiIII} absorption at
$v=-186$~{\kms} is the result of a blend, there is even more
flexibility in fitting the {\Lya} at $v<-130$~{\kms}.  In either case,
it is clear additional components are needed to fit the {\Lya}.

\begin{figure*}
\figurenum{3}
\epsscale{0.8}
\plotone{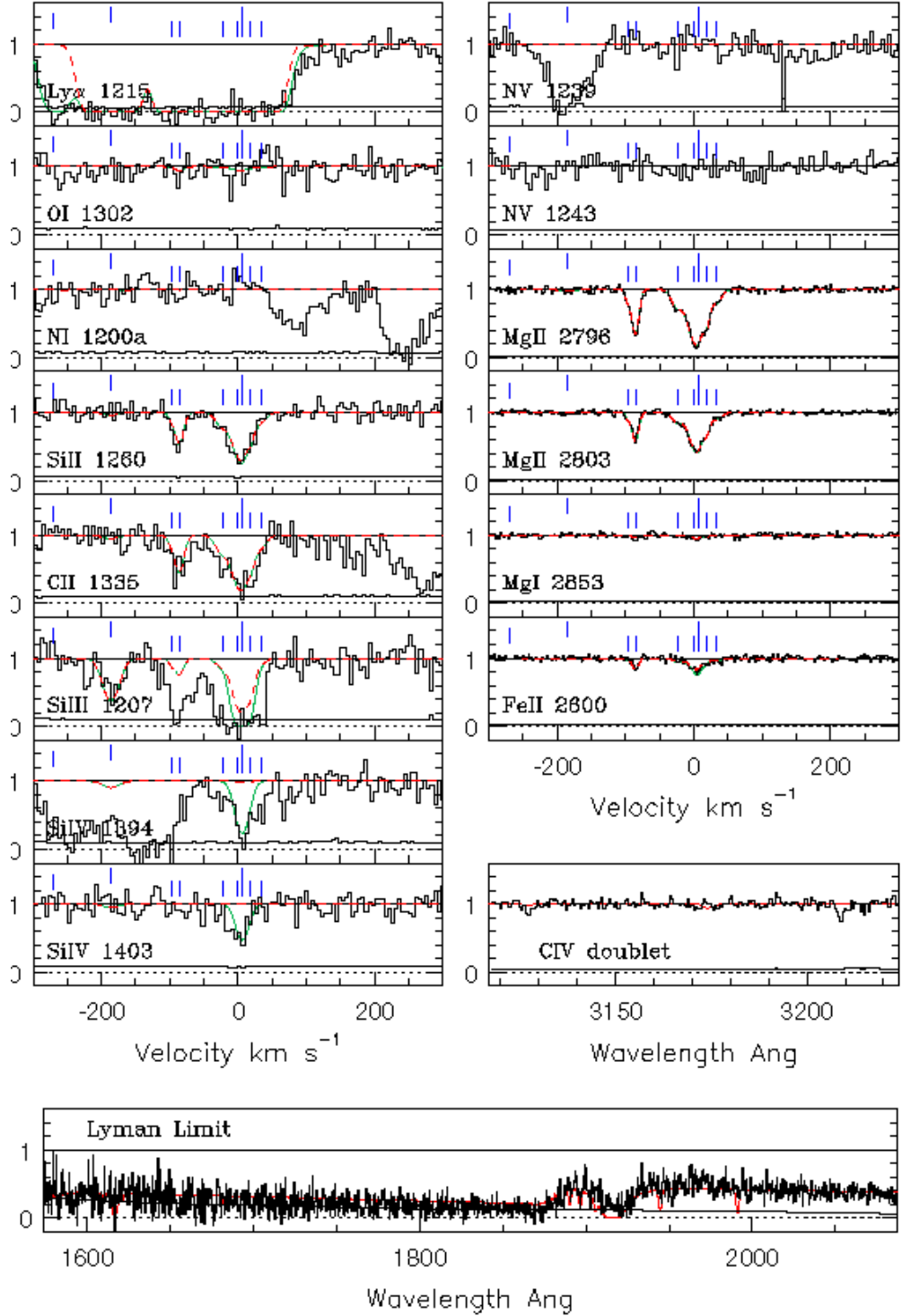}
\protect\caption{
\baselineskip = 0.7\baselineskip
The same as Fig~\ref{fig134}, except for the $z=1.0480$ system.
The {\CIV} was again covered in a ground based spectrum, but in
this case was not detected.
The lower row of ticks marks clouds {\MgII}$_1$--{\MgII}$_7$,
the middle row marks the one cloud {\Lya}$_1$, and the upper
row marks the two clouds {\SiIV}$_1$ and {\SiIII}$_1$ (see
Table~\ref{val104}).  The Lyman limit break at $\sim 1875$~{\AA}
is shown in the bottom panel, in the {\it HST}/FOS spectrum,
with model contributions from the two higher redshift systems
also included.
}
\label{fig104}
\end{figure*}

\begin{figure*}
\figurenum{4}
\epsscale{0.8}
\plotone{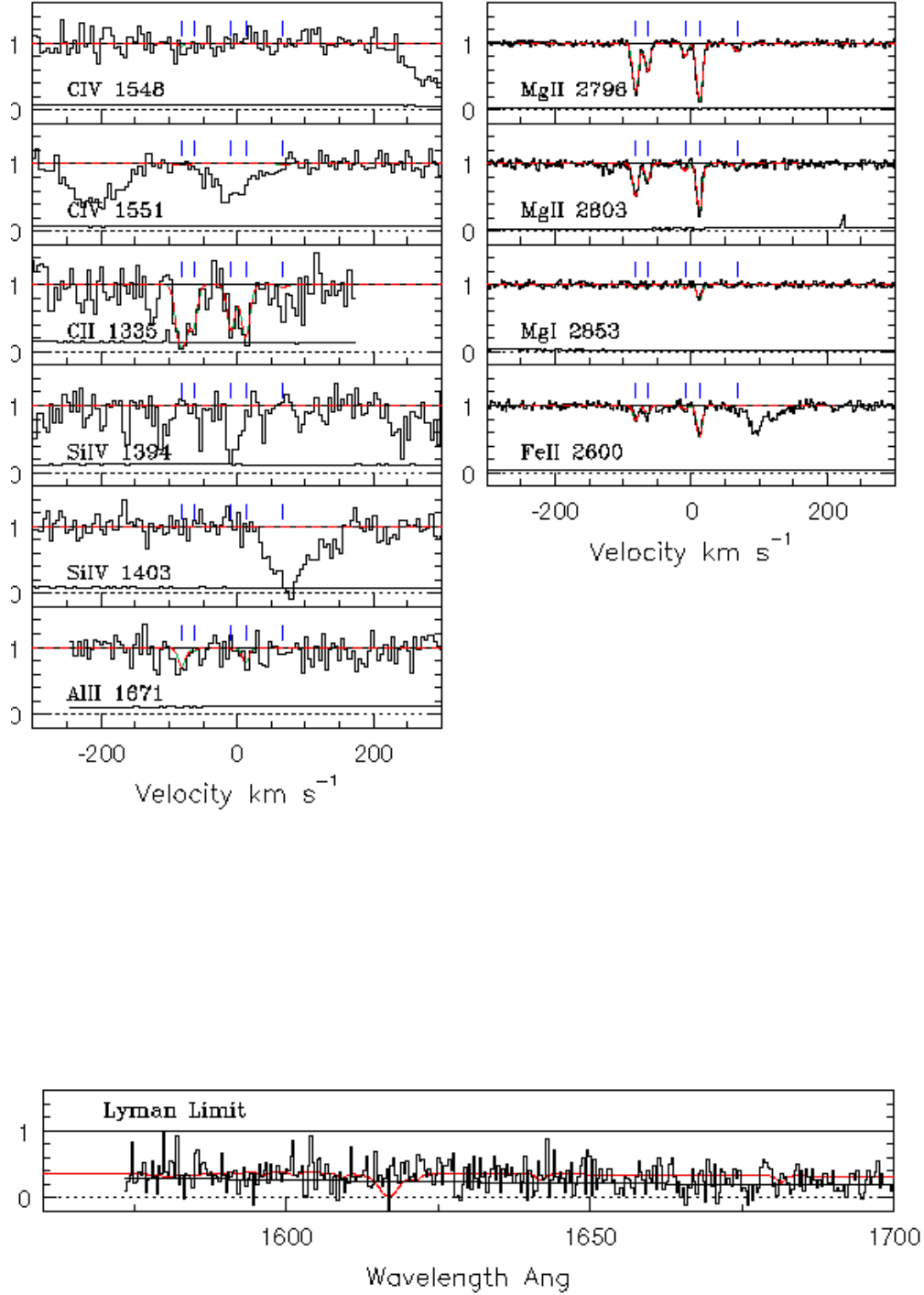}
\protect\caption{
\baselineskip = 0.7\baselineskip
The same as Fig~\ref{fig134}, except for the $z=0.7290$ system.  {\MgII}, {\MgI},
and {\FeII} are covered by the Keck I/HIRES spectrum, while other transitions
are covered by {\it HST}/STIS.
Here, the ticks mark the five low ionization clouds, {\MgII}$_1$--{\MgII}$_5$
in our model (see Table~\ref{val072}).  The lower panel shows the {\it HST}/FOS
spectrum in which the Lyman limit break for this system would be at $\sim 1576$~{\AA},
and is apparently not present at the blue edge of this spectrum.
}
\label{fig072} 
\end{figure*}

\begin{figure*}
\figurenum{5}
\epsscale{0.8}
\plotone{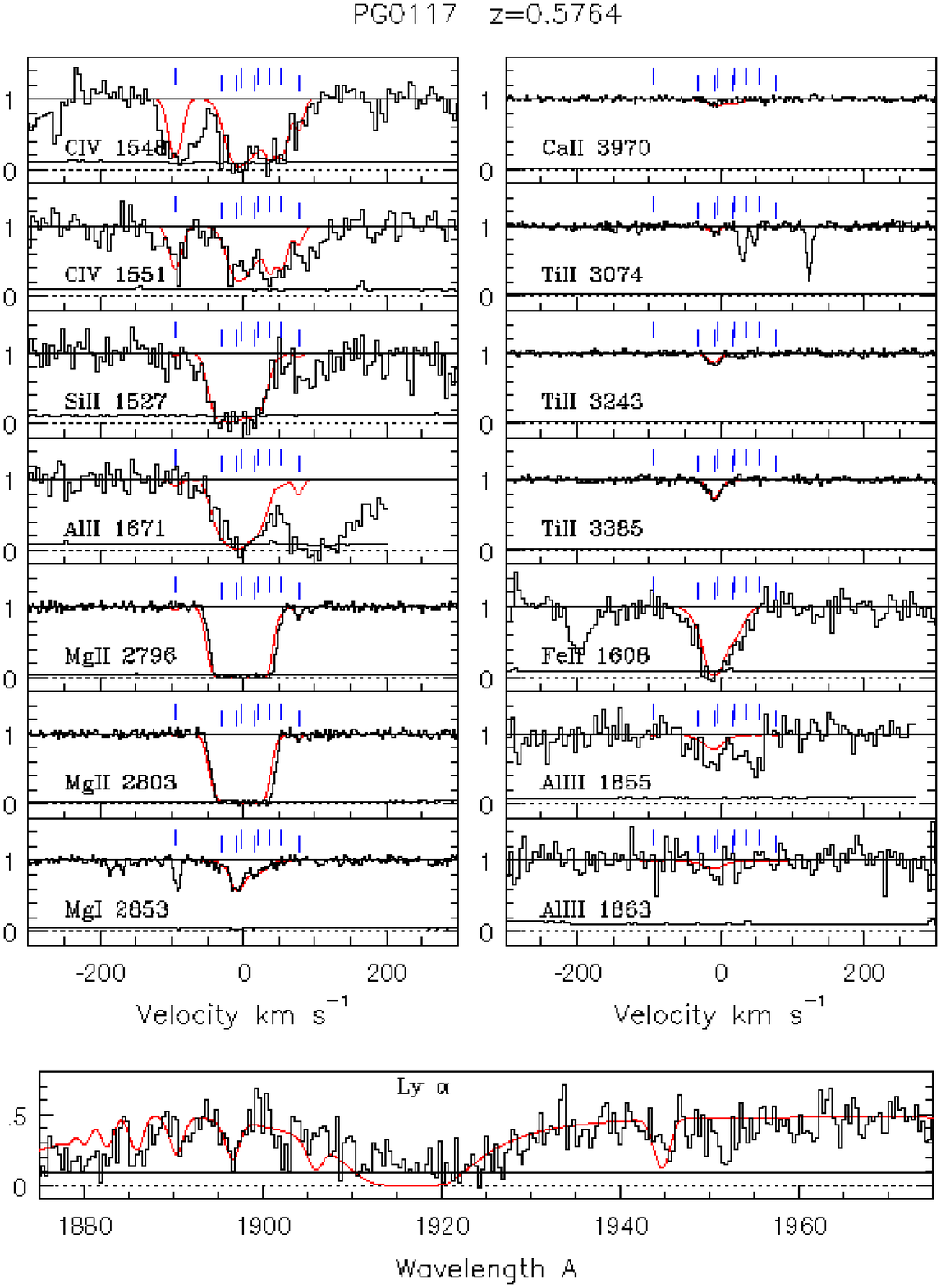}
\protect\caption{
\baselineskip = 0.7\baselineskip
The same as Fig~\ref{fig134}, except for the $z=0.5764$ system.  The
lower row of ticks marks positions of the very--low and low ionization
clouds, {\MgI}$_1$--{\MgI}$_3$, {\SiII}$_1$--{\SiII}$_3$, and
{\MgII}$_1$, which are superimposed on each other.  The upper row of
ticks marks the high ionization clouds, {\CIV}$_1$--{\CIV}$_5$.  The
parameters of these clouds are given in Table~\ref{val057}.  The
bottom panel shows coverage of {\Lya} for this system in a low
resolution {\it HST}/FOS spectrum, which was obtained in spectropolarimetry
mode.}
\label{fig057}
\end{figure*}

To simplify the process of constraining the metallicity of the clouds
in this system, all of the {\MgII} clouds ({\MgII}$_1$--{\MgII}$_7$)
were initially assumed to have the same value.  Based on the Lyman
limit break, which is dominated by the contribution of the {\HI} in
the {\MgII} clouds, the metallicity of these clouds was $\log
Z\sim-0.7$.  The metallicity of the {\SiIV} cloud ({\SiIV}$_1$) was
based on fitting the red wing of the {\Lya} profile. The {\Lya} in
this region, from $v\sim80$~{\kms} to $v\sim120$~{\kms} was clearly
underproduced by the {\MgII} clouds, shown as the dashed line in
Figure \ref{fig104}.  To fit the wing, the metallicity of the
collisionally ionized {\SiIV} cloud ({\SiIV}$_1$) was found to be
$\log Z\sim-1.4$.  For simplicity, in our tabulated model, also
displayed in Figure \ref{fig104}, the {\SiIII} and {\Lya} clouds
({\SiIII}$_1$ and {\Lya}$_1$) were taken to have the same metallicity
as this {\SiIV} cloud ({\SiIV}$_1$).

An alternative model is one in which the {\Lya} absorption is produced
primarily in a single, broad component.  This model assumes that the
apparent detected flux seen at $v\sim-250~{\kms}$ and $v\sim
-140~{\kms}$ is just the result of correlated noise in the data.  In
this case, the excess {\Lya} absorption, not produced by the {\MgII}
clouds, can be fit by a single cloud centered at $v\sim-98~{\kms}$,
with $\log N({\Lya}) \sim 16.3$~{\cmsq} and $b \sim 75~{\kms}$.
However, this cloud is constrained not to produce any detectable
absorption in any of the covered transitions. For any $\log U < -1.0$,
this would require extremely low metallicities, i.e. $\log Z < -2.7$.
Any broad, high ionization phase for this system is constrained in
this way as well.  It seems more likely that the {\Lya} absorption
arises in the several narrower clouds, as discussed above.

\subsubsection{Discussion}
\label{dis104}

The $z=1.0480$ system along this line of sight is very interesting, as
it would appear to be a classic strong {\MgII} system, with the
exception that it is {\CIV} deficient.  \citet{cwcCIV} proposed that
{\CIV} is produced by a corona phase around the host galaxy, based
upon the correlation between the equivalent width of the {\CIV} and
the velocity spread of the {\MgII} clouds.  In that model, the
outermost {\MgII} clouds were high velocity clouds associated to the
same star-forming processes that supported the corona.  Does our lack
of {\CIV} detection indicate that this galaxy is not undergoing star
formation?  There is also some evidence that {\CIV} deficient galaxies
tend to be reddish, which is indicative of gasless, older host
galaxies, perhaps elliptical \citep{archiveI}.  On the other hand, the
{\CIV} deficient system at $z=0.6600$ toward the quasar PG~$1317+274$
is likely to be related to a spiral galaxy \citep{steidel02}, although
it is a spiral with a reddish color ({\it B}$-${\it K}$=3.84$).
Recently, \citet{ding04} have suggested that {\CIV} deficiency is a
general feature of systems without a corona, and that there are a
variety of reasons for a lack of corona, meaning that many different
morphologies could cause {\CIV} deficient absorption.

For our system, one possible explanation for the {\CIV} deficiency is
that the host galaxy could be at a rather large impact parameter ($d
\ge 30~h^{-1}$~kpc).  Beyond the corona (which could be contained in
the inner disk), little or no {\CIV} would be observed.  Conversely,
if all the gas was in a particularly highly ionized corona, the {\CIV}
would be pushed into higher ionization states.  This would be
observable as very strong {\OVI} absorption, however {\OVI} was not
covered by our spectrum.  This gas would be constrained not to produce
any {\CIV} or {\NV} absorption.  This could be due to collisional
ionization, as is found in the $z=1.3430$ system, though at even
higher temperature, to limit the {\NV} and {\CIV} absorption.  Another
explanation could be a low metallicity corona, or a complete lack of a
corona.  The low metallicity case could be investigated with
observations of higher order transitions in the Lyman series.
Although the kinematics are suggestive of a spiral, in the absence of
direct information about the galaxy host, this could be an early type
galaxy.  A deeper image of the area around the quasar with
spectroscopic confirmation to detect the host of this system would be
very helpful in ruling out possibilities.  High resolution spectral
observations covering {\OVI} absorption for this system would also be
useful, though this would be a difficult observation because the flux
in that region is greatly reduced due to the Lyman limit breaks of the
$z=1.3430$ and $z=1.3250$ systems.

The {\Lya} absorption found in this system can be explained with
either a single, broad component, or with multiple narrower
components.  We tentatively favor a multiple component explanation
because of the alignment between the {\Lya} absorption and the
{\SiIII} components.  However, a very low metallicity, broad {\Lya}
component is possible.  Were this the case, this system would be
similar to the $z=0.9902$ system toward PG$1634+706$.  In that case, a
large, low metallicity halo was proposed to be responsible for most of
the {\Lya} absorption.  It is tempting to say that these are similar
systems, however either explanation is possible.

\subsection{The $z=0.7290$ System}
\subsubsection{Results}
\label{res072}

The $0.7290$ system is a multiple-cloud, weak {\MgII} absorber with a
total $W_r(2796)=0.240\pm0.008$~\AA.  In the HIRES spectrum {\MgII},
{\MgI}, and {\FeII} absorption are detected, and can be fit by five
distinct, narrow Voigt profiles ($3 < b < 5$~{\kms}), ranging in
velocity from $-82$~{\kms} to $67$~{\kms}, clouds
{\MgII}$_1$--{\MgII}$_5$ in Table~\ref{val072} \citep{cvc03}.
{\CII}~1335 is detected in the STIS spectrum, but {\AlII}~1671 cannot
be separated from the noise.  {\SiIV}~1403 is not detected, providing
a limit on {\SiIV} despite the features in the confused region of the
spectrum around {\SiIV}~1394.  Most importantly, {\CIVdblt} is not
detected, with $W_r(1548)<0.010$~\AA~ at a $3\sigma$ level.  For
$b=4$~{\kms}, this corresponds to $\log~N({\CIV}) = 12.4$~{\cmsq}.
The detected transitions, along with the region covering {\CIVdblt},
are shown in Figure~\ref{fig072}.  A Lyman limit break is not detected
for this system, however caution is in order because it would appear
in the first dozen pixels of the FOS spectrum, which may be too noisy
to be sure there is no feature there.

A single, low ionization phase was used to fit all the observed
transitions.  For simplicity, we assumed that all clouds had the same
metallicity.  Based on the lack of a Lyman limit break, $\log
N({\HI})<16.5$~{\cmsq} and a limit of $\log Z\ge0.6$ is placed on the
metallicity.  The ionization parameter can be constrained by the
{\FeII}~2600, assuming solar abundance pattern. Using a metallicity of
$\log Z\sim0.7$, clouds {\MgII}$_2$--{\MgII}$_4$, at $v=-64$, $-9$, and $12$~{\kms},
would have {\IPs} $-4.4<\log U<-4.1$.  Cloud {\MgII}$_1$, at $v=-82$~{\kms},
would have an {\IP} $-3.2<\log U<-2.9$.  Cloud {\MgII}$_5$, at $v=67$~{\kms}, is
only observed in {\MgII}, and so the {\IP} for it can only be
restricted to a lower limit of $\log U\le-2.5$, based on {\CIV}
production.  (The feature to the red of {\FeII}~2600 is {\MgI}~2853
from the $z=0.5764$ system.)  These values of $\log U$ are consistent
with the observed {\MgI}~2853 absorption, but they severely
underproduce {\CII}~1335.  An increase of the carbon abundance of
between $0.5$ and $2.0$~dex was required to bring the model into
agreement with the data.  Alternatively, we could increase $\log U$ to
bring the {\CII}~1335 into agreement, however, this would both
overproduce {\CIV} and require a large increase in the iron
abundance.

A $\log Z=0.7$ model for the low ionization phase is listed in Table
\ref{val072}.  The sizes of these weak clouds
({\MgII}$_1$--{\MgII}$_5$) are extremely small ($\lesssim1$~pc), and
they are quite cold due to their high metallicities.  To pass through
five such clouds along a single line of sight would imply a large
covering factor and/or an unusual geometry.  Because of these extreme
properties and because the determination of the lack of Lyman limit
break from the FOS data was so uncertain, we consider an alternative
model with $\log Z=-1.0$.  This model is problematic, because not only
is there a Lyman limit break created, but it also overproduces {\CIV}
and underproduced {\FeII}, even at the same ionization parameters as
the $\log Z=0.7$ model.  Should the carbon abundance adjustment be
decreased, overproduction of {\CIV} would still occur, and {\CII}
would then be underproduced.  We find it unlikely that this model is
correct.

The most striking thing about this system is the absence of detected
{\CIV}.  As just described, it is consistent with the low ionization
transitions to have negligible {\CIV} produced in the low ionization
phase, regardless of metallicity.  The lack of detected {\CIV} also
constrains the properties of any possible broad, high-ionization
phase.  Such a phase would have either extremely low metallicity or a
high ionization parameter.  The former could be diagnosed by high
resolution coverage of {\Lya} and the latter by {\NV} and {\OVI}
profiles.

\subsubsection{Discussion}
\label{dis072}

The $z=0.7290$ system is a multiple cloud weak {\MgII} system, but is
similar to the $z=1.0480$ strong {\MgII} system in the respect that it
has no detected {\CIV} absorption.  In this case, {\CIV} is covered by
the STIS spectrum, which has a much greater sensitivity than the
ground-based data used in the $z=1.0480$ system, and {\SiIV} is not
detected either.  This system has many of the same properties as the
$z=0.6600$ system toward CSO~$873$ presented by \citet{ding04}.
Again, we consider whether that the lack of the {\CIV} absorption is
due to galaxy structure, ionization, or low metallicity.

We inferred a very high metallicity for the low ionization phase.
This may suggest that a low metallicity corona is highly unlikely.
However, our high metallicity determination (${\log}Z\ge0.6$) is based
almost entirely on the lack of a Lyman limit break, which would only
be observed at the very end of the FOS spectrum.  This introduces
uncertainty into our conclusion.  If this high metallicity is correct,
though, it is very significant.  Single cloud weak {\MgII} clouds have
been shown to have metallicities greater than $0.1$ solar
\citep{weak2} and in many cases greater than solar \citep{weak1634},
but it has been suggested that multiple cloud weak {\MgII} systems are
completely different objects, with much lower metallicities
\citep{zonak,rosenberg}.  Also, single cloud weak systems commonly
have {\CIV} detected, in the redshift range of $0.5{\le}z{\le}1$
\citep{weak2}.  The single cloud weak {\MgII} absorbers are a
different class of absorbers than strong {\MgII} absorbers,
representing high metallicity regions that could fully account for the
{\Lya} forest with ${\log}N({\HI})>15.8$~{\cmsq}.  Our view of the
nature of the structures in which they arise could be greatly
influenced by the existence of multiple cloud versions of single cloud
weak absorbers.

A consequence of the high metallicity we have found is that these
clouds have low temperature.  This is not because they are dense, but
is a direct result of the high metallicity cooling the clouds, making
them very different from the {\MgI} phase of the $z=0.99$ system
toward PG~$1634+706$ presented by \citet{ding1634}.  In that system, a
very-low metallicity phase was required to explain the broad {\Lya}
profile, weak {\CIV}, and lack of detected {\NV} and {\OVI}.  The
authors suggested the phase could be a galactic halo or a diffuse
medium in an early type galaxy, but they had no direct galaxy
information.  In our case, we know that our absorption system is
associated with a red face-on barred spiral galaxy (approximately SBa)
at $36~h^{-1}$~kpc.  Without coverage of the {\Lya}, {\NV} or {\OVI},
we can only speculate that a similar broad component could exist, but
if it does we would know that an elliptical could produce such a
signature.  As with the $z=1.0480$ system, we note that several types
of {\CIV} deficient systems could exist, with the common element being
the lack of a traditional {\CIV} corona.

An oddity of this system is the abundance enhancement of carbon that
is required to fill out the observed {\CII} absorption.
\citet{abundance} presented results indicating that novae from
carbon-oxygen (CO) white dwarfs would produce excesses in carbon and
silicon.  It would be interesting to have {\SiII} coverage for this
system, to see if abundance increase is required for that as well.
This could indicate novae activity resulting from low-mass
($\lesssim1.2~M_\sun$) white dwarfs accreting and detonating.

\subsection{The $z=0.5764$ System}
\subsubsection{Results}
\label{res057}

This system is a very strong {\MgII} absorber, with
$W_r(2796)=0.902\pm0.002$~\AA~ \citep{cv01}.  In addition to the
{\MgII} doublet, {\MgI}, {\CaII}, and {\TiII} were detected in the
Keck I/HIRES spectrum.  These profiles could be fit with a minimum of
four Voigt profile components, with a velocity range of $-31$~{\kms}
to $77$~{\kms} \citep{cvc03} (see clouds {\MgI}$_1$--{\MgI}$_3$ and
{\MgII}$_1$ in Table \ref{val057}, which come from the fit).  In the {\it
HST}/STIS spectrum, {\CIVdblt}, {\SiII}~1527, {\AlII}~1671,
{\FeII}~1608, {\AlIII}~1855, and {\AlIII}~1863 were detected.  The
velocity range for the {\CIV} was greater than that for the lower
ionization gas, with a component at $-95$~{\kms}.  The detected
transitions from HIRES and STIS are shown in Figure~\ref{fig057}.  The
{\Lya} was covered only in a low resolution ($R\sim1300$) {\it
HST}/FOS spectrum, obtained in spectropolarimetry mode
\citep{koratkar}.  It gives only a rough measure of $W_r(1216)$ and a
tentative indication of the profile shape.  It is unclear whether the
profile indicates a damped {\Lya} absorber, or whether it can be
better fit with multiple components, with lower {\HI} column densities
\citep{rao}.

The {\MgIIdblt} and {\SiII}~1527 transitions are saturated and any
subcomponent structure cannot be discerned due to blending.  However,
the {\AlII}~1671 and the {\AlIIIdblt} transitions are not saturated,
and so provide a constraint on these transitions.  The very-low
ionization transitions, {\MgI}, {\FeII}, {\CaII}, and {\TiII}, also
show some kinematic structure.  Therefore, we use the Voigt profile
fit to the {\MgI}~2853, clouds {\MgI}$_1$--{\MgI}$_3$, as a starting
point for the photoionization models.  A very-low ionization phase,
with $-8.2 < \log U < -7.7$, can be used to fit the {\MgI} and {\CaII}
in cloud {\MgI}$_2$ at $v=-9$~{\kms}, but it underproduces {\TiII} and
{\FeII}.  To adequately fit the {\TiII} profile would require the
titanium abundance to be elevated by $\sim1.9$~dex relative to
magnesium and calcium.  Dust depletion is known to reduce titanium and
calcium absorption \citep{dust}.  Were this taken into account as
well, the abundance pattern adjustments would have to be even further
increased.  For the assumed range of {\IPs} and a metallicity of $\log
Z = -1.9$ (from the low ionization phase, as described below), the
cloud sizes range from $0.0008$ to $0.018$~pc.  For higher values of
the metallicity, the titanium abundance elevation would have to be
even more extreme.  The {\FeII} was adequately fit with an abundance
elevation of $\sim1.2$~dex.  Any variation in {\IP} or metallicity
would only serve to require a greater abundance pattern shift.
Figure~\ref{fig057} shows the contribution of the very-low ionization
phase (clouds {\MgI}$_1$--{\MgI}$_3$) to the {\MgII}, {\FeII}, and
{\SiII} as a dotted line, showing the need for a low ionization phase.

The low ionization phase was able to be constrained by the required
production of the saturated transitions, but more importantly by the
two {\AlIII} transitions.  For simplicity, we assumed three components
(clouds {\SiII}$_1$--{\SiII}$_3$) centered on the three very-low
ionization phase components, and found a plausible fit to the
{\SiII}~1527 profile with the minimum values of the column densities.
These components could only be constrained very loosely, using the
requirements that the {\MgII} profile be filled, and the {\AlII},
{\FeII}, and {\CaII} not be overproduced.  Attempting to fit the
components to the {\FeII} profile resulted in overproducing {\CaII} to
a greater degree than {\FeII} was underproduced by the very-low phase,
prompting us to vary the abundance of iron in that phase.

Because of saturation effects, specific values of constraints should
be viewed with caution.  For example, if the {\SiII}~1527 column
densities were larger, then $\log U$ would be constrained to be
higher.  Also, the model fits to the {\AlII}~1671 and {\MgIIdblt} are
not consistent with the data, suggesting additional abundance
parameter adjustments and/or a more complex phase structure.  It may
be possible to resolve this by adjusting the silicon abundance in the low
ionization phase, however.

Nonetheless, the essential conclusion concerning the low ionization
phase is that most of the {\CIV} absorption cannot be produced by
these same clouds.  This is not true for regions of velocity
coinciding with $v=-9$ and $v=16$~{\kms} clouds ({\SiII}$_2$ and
{\SiII}$_3$): these can be
adequately fit by the low ionization phase; when the {\IP} is
constrained by {\SiII} and {\AlIII}, the {\CIV} is also consistently
produced.

Cloud {\MgII}$_1$, at $v=77$~{\kms}, is able to produce {\MgII} and {\CIV} at that
velocity in a single phase having an {\IP} of $-2.3 < \log U < -2.2$.
However, an alternative model with two phases is also possible.  With
no additional unblended low ionization transitions, we will refrain
from speculating further on the existence of more than one phase in
this cloud.

Using the FOS data, and assuming that the observed {\Lya} profile was
indeed a single, damped {\Lya} (DLA) feature, we can consider the
metallicity of the very-low ionization clouds, {\MgI}$_1$--{\MgI}$_3$.  The metallicity
constraint is also subject to uncertainties due to saturation, but we
can consider our derived value as a lower limit, since we fit the
{\SiII}~1527 with the minimum column densities that were consistent
with its profile.
We found that the metallicity in the very-low ionization phase was
constrained to be $-2.0<\log Z<-1.8$, with a best fit at $\log
Z=-1.9$, which arises from the slight variations in absorption
strengths between transitions caused by varying metallicity.  This
effect happens in the optically thick regime.  

Due to the narrow range of the velocity distribution of the low
ionization gas, we were unable to model the {\Lya} profile as multiple
individual components, as suggested in \citet{rao}.  It seems more
likely that the {\Lya} is dominated by a single cloud, or by a few
clouds very close together in velocity space.

A higher ionization phase is needed to produce the {\CIV} absorption
in this system, for the main reason that the {\CIVdblt} profiles have
absorption lying at both higher and lower velocities than the low
ionization profiles.  Since {\CIV} is the only high ionization
transition covered, the only constraint on its ionization parameter is
the non-detection of low ionization transitions at some velocities.
This restricts the {\IP} to $\log U\ge-1.8$.  A plausible model,
including clouds {\CIV}$_1$--{\CIV}$_5$ with $\log U = -1.8$ is
summarized in Table \ref{val057}.

\subsubsection{Discussion}
\label{dis057}

This DLA system was difficult to model, due to the saturation
in the {\MgII} transitions, and the complex structure in the {\MgI}
transition.  It is unlikely that our solution is unique, but we
are still able to draw general conclusions from our chosen model and
our other failed possibilities.  It is clear that two separate
ionization phases are required to produce the observed {\MgI}, {\MgII},
and {\SiII}.  Even with these extra phases, large abundance
pattern variations are needed to account for the {\TiII} absorption.

Having an extra very-low ionization phase is consistent with other
observations of DLAs.  DLA systems have been observed to have
molecules \citep{ge,ledoux02,ledoux03}, which indicates a very cold,
high density environment, consistent with a small, very-low ionization
region.  Narrow velocity widths would be expected from such a cold
phase of gas.  \citet{radioDLA} present a DLA system with a very
narrow velocity width in one of the components of the $21~$cm
absorption, consistent with the Doppler parameters in our models of
the very low ionization phase of the $z=0.5764$ system.  The sizes
that we have estimated from our models are as small as $\sim150$~AU,
which at first may seem remarkable.  However, it is known, from a
variety of observational techniques, that there exists significant
structure down to scales of $100$~AU in the interstellar medium of
the Milky Way \citep{andrews01,meyer99,watson96,meyer96,frail94}.
It is interesting to note that, in the absence of pressure balance,
there is still no clear explanation for the small scale clumping in the
Milky Way \citep{elmegreen,heiles,walker}.  In terms of small scale
structure in extragalactic absorbers, \citet{ding1634} found similar
ones in the strong {\MgII} absorber at $z=1.0414$ in the
PG~$1634+706$ line of sight.

The {\CIV} absorber in this system is weaker relative to the {\MgII}
than is expected for a classic, strong {\MgII} system
(\cite{archiveII}; although those authors were working with a very
small sample, which should be taken into consideration).  The {\CIV}
does not appear to be associated with lower ionization gas, even just
based on the stark velocity differences between the observed
absorption in each.  This is consistent with the DLA absorption being
due to a contained region within a larger structure.

If the line of sight did not happen to pass through this contained
region, the system would have the same absorption properties as an
ordinary {\MgII} absorber.  This idea is supported by the fact that
the candidate galaxy believed to be associated with this absorber is
at a very small impact parameter ($d=5.1~h^{-1}$~kpc).  It is
interesting that the {\CIV} absorption in this DLA is weaker than any
other DLA's in the small sample of \citet{archiveII}, especially
because this is associated with a potential elliptical galaxy.  In
some cases the {\CIV} absorption in a DLA could be enhanced by the
contributions from very high column density {\MgII} clouds, however,
in this case the effect of this enhancement is not large.

\section{System Summary}
\label{sum}

We present the results from our photoionization and collisional
ionization modeling of five intervening quasar absorption line systems
along the line of sight toward the quasar PG~$0117+213$.  Starting
with the highest redshift system, we analyzed the absorption features
of the $z=1.3430$, $z=1.3250$, $z=1.0480$, $z=0.7290$, and $z=0.5764$
absorption systems.

\begin{itemize}
\item The three phase $z=1.3430$ system is a multiple cloud weak {\MgII}
      absorber which had low ionization phase properties similar to
      many strong {\MgII} absorbers.  The high ionization components
      were found to be unrelated to the low ionization gas due to the
      differences in their kinematics. In addition, both high
      ionization components are found to be collisionally ionized,
      with different metallicities, one of which varies significantly
      from the metallicity of the low ionization gas.  A broad, low
      metallicity {\Lya} component was needed to fill in the wings of
      the {\Lya} absorption.  The nature of the host galaxy is not
      known, but we suggest that this system could be produced in the
      outskirts of a spiral disk, or in a dwarf galaxy.

\item The three phase $z=1.3250$ system is a multiple cloud weak {\MgII}
      absorber as well, but its components are spread out in a
      distribution similar to what we might expect from an elliptical
      galaxy, or a spiral with no disk gas contribution
      \citep{kinmod}.  High ionization {\OVI} gas was seen over the
      same range as the low ionization gas, but again it was found not
      to be related to the lower ionization transitions.  This system
      is strong in {\CIV} absorption, but has very weak {\NV}
      absorption, requiring a significant nitrogen abundance pattern
      shift from solar.  This system kinematically suggests an
      elliptical galaxy or lack of disk gas component, but it would be
      an exception to the possible tendency for red galaxies to be
      {\CIV} deficient \citep{archiveII}.  On the other hand, some
      lines of sight through spiral galaxies in a statistical sample
      (those with weak absorption contributions from the disk) would
      have kinematics consistent with the $z=1.3250$ system as
      well \citep{kinmod}.

\item The four phase $z=1.0480$ system is a {\CIV} deficient strong
      {\MgII} system.  With only a potential host candidate in the
      field, we can only say that the kinematics are consistent with a
      spiral galaxy.  However, the {\CIV} deficiency suggests a highly
      ionized or sparse corona, or even the absence of a corona, such
      as is expected for an elliptical galaxy.

\item The single phase $z=0.7290$ system is a multiple cloud weak
      {\MgII} system that has components kinematically spread out as
      could be expected for an elliptical galaxy.  However, the galaxy
      host is known to be a very red, barred spiral galaxy in the
      optical images of the field.  This is still consistent with the
      fact that the absorber is {\CIV} deficient, suggesting that the
      corona is weak or absent.  This absorber may be a typical {\CIV}
      deficient system, although we found it to possibly have
      supersolar metallicity with unusually small, low ionization
      cloud sizes.

\item The three phase $z=0.5764$ system is a damped {\Lya} absorber,
      which is possibly produced by a candidate galaxy at a small
      impact parameter ($5.1~h^{-1}$~kpc), which is very red ({\it
      B-K}$=4.00$).  A separate very-low ionization phase is
      required to produce the observed absorption in the very-low
      ionization species.  This system required an abundance pattern
      adjustment to account for all low ionization species observed
      ({\CaII}, {\MgII}, {\FeII}, and {\TiII}).  {\CIV} absorption is
      observed from this galaxy, but some of it cannot be related to
      the lower ionization gas due to its non-coincident velocity
      range.  This is consistent with the idea that a DLA has low
      ionization absorption arising from a small region within a
      larger, somewhat unrelated structure.

\end{itemize}

\section{Conclusion}
\label{conc}

We think of the Milky Way Galaxy as a typical spiral galaxy, with a
structured multi--phase interstellar medium.  The Milky Way also has a
highly ionized corona, a few kiloparsecs thick around the disk, which
produces broad absorption features in {\CIV}, {\NV}, and {\OVI}
(\cite{kp15}, and references therein).  These broad features do often
have component structure, but the structure is usually restricted to
the same velocity range as that of the low ionization absorption.
Most giant galaxies in the nearby universe are spirals, and there is
also an absorption cross section presented by early type galaxies.
Furthermore, there is a non--negligible cross section of dwarf
galaxies, which presumably produce some absorption.  There are five
{\MgII} absorption systems along the quasar PG~$0117+213$ line of
sight, which are presumably produced by a random, but small, sampling
of galaxies with redshifts ranging from $z=0.5764$ to $z=1.3430$.  We
conclude with a discussion of how these five systems compare to the
absorption seen looking through the Milky Way, and how they may
resemble or differ from other kinds of absorbing structures that we
see in the nearby universe.

Three of the {\MgII} absorbers in this study are classified as
multiple cloud weak {\MgII} absorption systems.  These absorbers are
about two-thirds as common as strong {\MgII} absorbers \citep{weak1}.
The cross section presented by their hosts must then be significant,
about two-thirds of that presented by the luminous galaxies that
produce the strong {\MgII} absorption.  Because the equivalent width
division at $0.3~$\AA~ is arbitrary, we would expect some of these
multiple cloud weak {\MgII} systems are simply less extreme versions
of strong {\MgII} absorbers, which are associated with giant galaxies.
Indeed, the $z=0.7290$ system toward PG~$0117+213$ is found to be
associated with a $\sim3~L_K^\star$ face on SBa galaxy at an impact
parameter of $36~h^{-1}~$kpc \citep{csv96}.  This distance is close to
the boundary of
$38~h^{-1}\left(\frac{L_K}{L_K^\star}\right)^{0.15}$~kpc, within which
strong {\MgII} absorption should occur \citep{steidel95}.  This
absorber is a very red galaxy, and a {\CIV} deficient system as well
(which \citet{archiveI} suggests is an indicator of early type
galaxies, low star formation rate, or a long time since the last
episode of star formation).

The other two multiple cloud weak {\MgII} absorbers at $z=1.3430$ and
$z=1.3250$ are presently too distant for the galaxy properties to be
directly explored.  The $z=1.3250$ system has kinematics similar to
the $z=0.7290$ system, tempting us to say that the former is a very
red galaxy, like the latter, and possibly an early type galaxy.
However, the $z=1.3250$ system also has strong {\OVI} and {\CIV}
absorption, loosely covering the same range as the low ionization
absorption, and it is not {\CIV} deficient.  Perhaps some selected
lines of sight through the Milky Way disk would look like this. 

The $z=1.3430$ system is different from the $z=1.3250$ system in that
it has a kinematically dominant component in the low ionization
absorption, perhaps suggestive of a disk.  Additionally, the
$z=1.3430$ system has broad absorption in the high ionization gas that
is quite kinematically unrelated to the low ionization gas.
\citet{zonak} have discussed the possible absorption signature of
dwarf galaxies and their winds, and their possible relation to
multiple cloud weak {\MgII} systems.  Dwarf galaxies must present some
significant absorption cross section, and multiple cloud weak {\MgII}
absorbers could be their signature in some cases.  Both the $z=1.3250$
and the $z=1.3430$ systems could fall in this category.
Generally, the three multiple cloud weak {\MgII} absorbers along this
PG~$0117+213$ line of sight illustrate the idea that there is likely to
be more than one type of origin for this class of system.

The absorption from the $z=0.5764$ DLA system could be consistent with
what would be expected from some lines of sight through the Milky Way
Galaxy.  However, this system is believed to be related to a very red
galaxy at a relatively low impact parameter.  The {\CIV} absorption
appears to be unrelated to the gas producing the DLA, but it is still
substantial for a very red galaxy, many of which are {\CIV} deficient
(for example, our $z=0.7290$ system).  On the other hand, the {\CIV}
absorption is relatively small compared to that produced by other DLAs
\citep{archiveII}.  It will be important to analyze the kinematics of
{\CIV} relative to {\MgII} as the sample of DLA systems studied at
high resolution in multiple transitions grows.

The $z=1.0480$ system could be a fairly standard case of a strong
{\MgII} absorber that is {\CIV} deficient. The possible candidate
galaxy has an impact parameter of $22.7~h^{-1}$~kpc, which is well
within the boundary of
$38~h^{-1}\left(\frac{L_K}{L_K^\star}\right)^{0.15}$~kpc for strong
{\MgII} absorption. This could imply that this system is an early type
galaxy (elliptical or spiral with little star formation), though this
is by no means definite.

As it turns out, none of the five PG~$0117+213$ {\MgII} absorbers give
the absorption signature expected for a combination of a disk, gaseous
corona, and occasional high velocity clouds like we see from
the Milky Way Galaxy.  However, not all absorption systems at
$z\sim1$ are so different from the expected signature of a spiral
galaxy.  In a couple of strong {\MgII} absorbers (e.g. PG~$1206+459$
at $z=0.9276$ \citep{ding1206} and PG~$1248+401$ at $z=0.7729$
\citep{ding04}), the absorption features in high ionization gas do
resemble coronae that kinematically encompass the dominant low
ionization features.  In some of the PG~$0117+213$ systems studied
here, and in some other systems at $z\sim1$, it also seems that "high
velocity clouds" are producing both low and high ionization absorption
\citep{archiveII}.  It will be of interest as data sets grow to study
whether there is an evolution of the properties of high velocity clouds.

High resolution coverage of a variety of transitions for the five
PG~$0117+213$ absorption systems provided information about several
classes of absorbers: multiple-cloud weak {\MgII}, {\CIV}-deficient,
and damped {\Lya}.  In addition to these classes, the {\MgII} absorber
population at $z=1$ includes single-cloud weak {\MgII} absorbers and
traditional spiral galaxies with disks and coronae.  Although they
share some elements in common, we have seen that there is substantial
variation in the absorption properties within a class.  We plan a
study of a larger sample of $\sim100$ {\MgII} absorbers covering many
key transitions with high resolution is feasible, using archival {\it
HST}/STIS data in conjunction with ground-based data, when available.
With a sample of that size, it will be possible to have $>10$ objects within
each class, which will allow separation of some classes into
sub-classes, and more specific connections of other classes to physical
processes and/or to morphologies of luminous hosts.

The authors would like to thank Buell Jannuzi for providing them with
an early release of the STIS spectrum, and Sandhya Rao and Anuradha Koratkar
for help with interpretation of the FOS spectrum.  Special thanks to Michele
Crowl for her careful proofreading.  This research was funded by NASA
under grants NAG 5-6399, NNG04GE73G, and HST-GO-08672.01-A, the latter
from the Space Telescope Science Institute, which is operated by AURA,
Inc., under NASA contract NAS 5-26555; and by NSF under grant
AST-04-07138.  JRM was partially funded by the NSF REU program.



\newpage

\newpage

\newpage

\newpage

\newpage

%
%

\newpage
\begin{deluxetable}{lrr}
\tablenum{1}
\tabletypesize{\footnotesize}
\tablewidth{0pt}
\tablecaption{Rest Frame Equivalent Widths for $z=1.3430$}
\tablehead{
\colhead{Transition}   &
\colhead{EW (\AA)}   &
\colhead{$\sigma_{EW}$} 
}

\startdata
{\Lya} & $1.14$ & $0.01$\\
{\HI}~$1206$ & $0.76$ & $0.01$\\
{\SiII}~$1206$ & $0.144$ & $0.008$\\
{\SiII}~$1190$ & $0.033$ & $0.003$\\
{\CII}~$1036$ & $0.058$ & $0.004$\\
{\NII}~$1084$ & $0.050$ & $0.004$\\
{\CII}~$977$ & $0.69$ & $0.01$\\
{\NIII}~$998$ & $0.164$ & $0.009$\\
{\NV}~$1239$ & $0.75$ & $0.02$\\
{\NV}~$1243$ & $0.23$ & $0.01$\\
{\OVI}~$1032$ & $0.37$ & $0.01$\\
{\OVI}~$1038$ & $0.27$ & $0.01$\\
{\MgII}~$2803$ & $0.147$ & $0.004$\\
{\FeII}~$2587$ & $0.013$ & $0.001$\\
{\CIVdblt} & $0.67$ & $0.02$\\

\hline
\enddata
\vglue -0.05in
\tablecomments{
\baselineskip=0.7\baselineskip
This table shows the equivalent widths for the transitions shown in
Fig~\ref{fig134}.  Included are the $\sigma$ values for each
measurement, or $3~\sigma$ limits for non-detections.  Blends with
other absorption features that prohibit equivalent width measurements
are marked appropriately.
}
\label{ew134}
\end{deluxetable}
\clearpage

\newpage
\begin{deluxetable}{lrr}
\tablenum{2}
\tabletypesize{\footnotesize}
\tablewidth{0pt}
\tablecaption{Rest Frame Equivalent Widths for $z=1.3250$}
\tablehead{
\colhead{Transition}   &
\colhead{EW (\AA)}   &
\colhead{$\sigma_{EW}$} 
}

\startdata
{\Lya} & $1.41$ & $0.01$\\
{\HI}~$1206$ & $1.67$ & $0.02$\\
{\SiII}~$1260$ & $0.08$ & $0.02$\\
{\CII}~$1335$ & $0.065$ & $0.008$\\
{\SiIII}~$1207$ & blend & n/a\\
{\NV}~$1239$ & $0.078$ & $0.007$\\
{\NV}~$1243$ & $<0.012$ & ...\\
{\OVI}~$1032$ & $0.82$ & $0.01$\\
{\OVI}~$1038$ & $0.34$ & $0.01$\\
{\MgII}~$2796$ & $0.312$ & $0.007$\\
{\MgII}~$2803$ & $0.191$ & $0.007$\\
{\MgI}~$2853$ & $0.006$ & $0.002$\\
{\FeII}~$2600$ & $0.035$ & $0.003$\\
{\CIVdblt} & $0.89$ & $0.03$\\

\hline
\enddata
\vglue -0.05in
\tablecomments{
\baselineskip=0.7\baselineskip
Same as Table \ref{ew134}, except for the transitions shown in
Fig~\ref{fig132}. 
}
\label{ew132}
\end{deluxetable}
\clearpage

\newpage
\begin{deluxetable}{lrr}
\tablenum{3}
\tabletypesize{\footnotesize}
\tablewidth{0pt}
\tablecaption{Rest Frame Equivalent Widths for $z=1.0480$}
\tablehead{
\colhead{Transition}   &
\colhead{EW (\AA)}   &
\colhead{$\sigma_{EW}$} 
}

\startdata
{\Lya} & $1.58$ & $0.01$\\
{\OI}~$1302$ & $<0.010$ & ...\\
{\NI}~$1200a$ & $<0.008$ & ...\\
{\SiII}~$1260$ & $0.128$ & $0.005$\\
{\CII}~$1335$ & $0.28$ & $0.01$\\
{\SiIII}~$1207$ & $0.47$ & $0.01$\\
{\SiIV}~$1394$ & $0.19$ & $0.01$\\
{\SiIV}~$1403$ & $0.085$ & $0.007$\\
{\NV}~$1239$ & $<0.009$ & ...\\
{\NV}~$1243$ & $<0.007$ & ...\\
{\MgII}~$2796$ & $0.417$ & $0.005$\\
{\MgII}~$2803$ & $0.257$ & $0.005$\\
{\MgI}~$2853$ & $0.026$ & $0.004$\\
{\FeII}~$2600$ & $0.073$ & $0.004$\\
{\CIVdblt} & $<0.08$ & ...\\

\hline
\enddata
\vglue -0.05in
\tablecomments{
\baselineskip=0.7\baselineskip
Same as Table \ref{ew134}, except for the transitions shown in
Fig~\ref{fig104}. 
}
\label{ew104}
\end{deluxetable}
\clearpage

\newpage
\begin{deluxetable}{lrr}
\tablenum{4}
\tabletypesize{\footnotesize}
\tablewidth{0pt}
\tablecaption{Rest Frame Equivalent Widths for $z=0.7290$}
\tablehead{
\colhead{Transition}   &
\colhead{EW (\AA)}   &
\colhead{$\sigma_{EW}$} 
}

\startdata
{\CIV}~$1548$ & $<0.010$ & ...\\
{\CIV}~$1551$ & blend & n/a\\
{\CII}~$1335$ & $0.22$ & $0.02$\\
{\SiIV}~$1394$ & blend & n/a\\
{\SiIV}~$1403$ & $<0.010$ & ...\\
{\AlII}~$1671$ & $0.043$ & $0.007$\\
{\MgII}~$2796$ & $0.240$ & $0.008$\\
{\MgII}~$2803$ & $0.137$ & $0.005$\\
{\MgI}~$2853$ & $0.015$ & $0.001$\\
{\FeII}~$2600$ & $0.058$ & $0.005$\\

\hline
\enddata
\vglue -0.05in
\tablecomments{
\baselineskip=0.7\baselineskip
Same as Table \ref{ew134}, except for the transitions shown in
Fig~\ref{fig072}. 
}
\label{ew072}
\end{deluxetable}
\clearpage

\newpage
\begin{deluxetable}{lrr}
\tablenum{5}
\tabletypesize{\footnotesize}
\tablewidth{0pt}
\tablecaption{Rest Frame Equivalent Widths for $z=0.5764$}
\tablehead{
\colhead{Transition}   &
\colhead{EW (\AA)}   &
\colhead{$\sigma_{EW}$} 
}

\startdata
{\Lya} & $11.2$ & $1.1$\\
{\CIV}~$1548$ & $0.76$ & $0.02$\\
{\CIV}~$1551$ & $0.52$ & $0.02$\\
{\SiII}~$1527$ & $0.55$ & $0.02$\\
{\AlII}~$1671$ & blend & n/a\\
{\FeII}~$1608$ & $0.34$ & $0.01$\\
{\AlIII}~$1855$ & $0.22$ & $0.01$\\
{\AlIII}~$1863$ & $<0.03$ & ...\\
{\MgII}~$2796$ & $0.902$ & $0.002$\\
{\MgII}~$2803$ & $0.841$ & $0.006$\\
{\MgI}~$2853$ & $0.149$ & $0.006$\\
{\CaII}~$3970$ & $0.018$ & $0.003$\\
{\TiII}~$3074$ & $0.012$ & $0.002$\\
{\TiII}~$3243$ & $0.055$ & $0.004$\\
{\TiII}~$3385$ & $0.073$ & $0.003$\\

\hline
\enddata
\vglue -0.05in
\tablecomments{
\baselineskip=0.7\baselineskip
Same as Table \ref{ew134}, except for the transitions shown in
Fig~\ref{fig057}. Measurement of {\Lya} is from a low resolution
{\it HST}/FOS spectrum.
}
\label{ew057}
\end{deluxetable}
\clearpage

%
%

\newpage
\begin{deluxetable}{cccccccccccccccc}
\tablenum{6}
\tabletypesize{\scriptsize}
\rotate
\tablewidth{0pt}
\tablecaption{Model Parameters for $z=1.3430$}
\tablehead{
\colhead{Cloud}   &
\colhead{v}   &
\colhead{$\log{{\rm Z}\over{Z_{\sun}}}$}   &
\colhead{$\log~$U} &
\colhead{T} &
\colhead{Size} &
\colhead{n$_H$} &
\colhead{$\log$}   &
\colhead{$\log$}   &
\colhead{$\log$}   &
\colhead{$\log$}   &
\colhead{$\log$}   &
\colhead{$\log$}   &
\colhead{b$_{\HI}$} &
\colhead{b$_{opt}$} \\
\colhead{}   &
\colhead{{\kms}}   &
\colhead{}   &
\colhead{} &
\colhead{K} &
\colhead{pc} &
\colhead{{\cc}} &
\colhead{N$_{Tot}$(H)}   &
\colhead{N({\HI})}   & 
\colhead{N({\MgII})}   &
\colhead{N({\CIV})}   &
\colhead{N({\NV})}   & 
\colhead{N({\OVI})}   &
\colhead{{\kms}}   &
\colhead{{\kms}} 
}

\startdata
{\MgII}$_1$ & $-137$ & $-0.3$ & $-2.8$ & $9000$ & $11$ & $0.009$ & $17.5$ & $15.1$ & $11.8$ & $12.1$ & $10.3$ & $9.2$ & $12.7$ & $4.1$\\
{\MgII}$_2$ & $-104$ & $-0.3$ & $-3.3$ & $9000$ & $2.1$ & $0.028$ & $17.2$ & $15.4$ & $12.0$ & $11.0$ & $9.2$ & $0.0$ & $13.5$ & $6.4$\\
{\MgII}$_3$ & $-59$ & $-0.3$ & $-2.8$ & $9000$ & $6.8$ & $0.009$ & $17.3$ & $14.9$ & $11.6$ & $11.9$ & $10.1$ & $9.0$ & $14.2$ & $7.6$\\
{\MgII}$_4$ & $1$ & $-0.3$ & $-4.0$ & $8000$ & $1.7$ & $0.14$ & $17.8$ & $16.6$ & $13.0$ & $10.1$ & $7.7$ & $0.0$ & $11.5$ & $4.0$\\
{\MgII}$_5$ & $11$ & $-0.3$ & $-2.8$ & $9000$ & $11$ & $0.009$ & $17.5$ & $15.1$ & $11.8$ & $12.1$ & $10.3$ & $9.2$ & $12.6$ & $3.8$\\
{\OVI}$_1$ & $-44$ & $-0.2$ & $...$ & $210000$ & $1.0$ & $...$ & $19.2$ & $13.6$ & $9.4$ & $13.5$ & $14.1$ & $14.5$ & $72.9$ & $45.0$\\
{\CIV}$_1$ & $-147$ & $1.0$ & $...$ & $130000$ & $0.44$ & $...$ & $18.4$ & $13.3$ & $10.7$ & $15.0$ & $14.0$ & $11.7$ & $53.7$ & $30.0$\\
{\Lya}$_1$ & $-73$ & $-2.5$ & $-2.2$ & $24000$ & $20000$ & $0.002$ & $20.1$ & $16.8$ & $11.3$ & $13.1$ & $11.5$ & $11.0$ & $45.0$ & ... \\


\hline
\enddata
\vglue -0.05in
\tablecomments{
\baselineskip=0.7\baselineskip
Model parameters for the system at $z=1.3430$.  Results of this model
are displayed superimposed on the data in Fig \ref{fig134}.  A ``...''
entry in the {\IP} and $n_H$ columns indicates a collisionally ionized cloud.
All column densities are expressed in units of {\cmsq}.  $b_{opt}$ is
the Doppler parameter for the transition that was ``optimized on'' in
modeling, the same transition as listed in the cloud designation column.  The ionization fraction, $f$, can be determinted from $\log f = \log {\rm N}({\HI}) - \log {\rm N}_{Tot}$(H).}
\label{val134}
\end{deluxetable}
\clearpage

\newpage
\begin{deluxetable}{ccccccccccccccccc}
\tablenum{7}
\tabletypesize{\scriptsize}
\rotate
\tablewidth{0pt}
\tablecaption{Model Parameters for $z=1.3250$}
\tablehead{
\colhead{Cloud}   &
\colhead{v}   &
\colhead{$\log{{\rm Z}\over{Z_{\sun}}}$}   &
\colhead{$\log~$U} &
\colhead{T} &
\colhead{Size} &
\colhead{n$_H$} &
\colhead{$\log$}   &
\colhead{$\log$}   &
\colhead{$\log$}   &
\colhead{$\log$}   &
\colhead{$\log$}   &
\colhead{$\log$}   &
\colhead{b$_{\HI}$} &
\colhead{b$_{opt}$} \\
\colhead{}   &
\colhead{{\kms}}   &
\colhead{}   &
\colhead{} &
\colhead{K} &
\colhead{pc} &
\colhead{{\cc}} &
\colhead{N$_{Tot}$(H)}   &
\colhead{N({\HI})}   &
\colhead{N({\MgII})}   &
\colhead{N({\MgI})}   &
\colhead{N({\CIV})}   &
\colhead{N({\OVI})}   &
\colhead{{\kms}}   &
\colhead{{\kms}}   
}

\startdata
{\MgII}$_1$ & $-82$ & $-0.3$ & $-3.5$ & $9000$ & $0.40$ & $0.043$ & $16.7$ & $15.0$ & $11.6$ & $9.4$ & $10.1$ & $0.0$ & $6.4$ & $1.3$\\
{\MgII}$_2$ & $-60$ & $-0.3$ & $-3.3$ & $9000$ & $10$ & $0.027$ & $17.9$ & $16.0$ & $12.7$ & $10.3$ & $11.7$ & $0.0$ & $12.1$ & $2.8$\\
{\MgII}$_3$ & $-35$ & $-0.3$ & $-2.4$ & $10000$ & $19$ & $0.0034$ & $17.3$ & $14.5$ & $11.2$ & $8.3$ & $12.5$ & $10.1$ & $6.9$ & $1.4$\\
{\MgII}$_4$ & $-5$ & $-0.3$ & $-3.5$ & $9000$ & $2.0$ & $0.043$ & $17.4$ & $15.7$ & $12.3$ & $10.1$ & $10.8$ & $0.0$ & $12.1$ & $2.8$\\
{\MgII}$_5$ & $48$ & $-0.3$ & $-3.6$ & $9000$ & $2.7$ & $0.054$ & $17.6$ & $16.0$ & $12.6$ & $8.9$ & $10.8$ & $0.0$ & $12.1$ & $3.3$\\
{\MgII}$_6$ & $143$ & $-0.3$ & $-2.2$ & $11000$ & $640$ & $0.002$ & $18.6$ & $15.6$ & $12.3$ & $9.2$ & $14.0$ & $12.0$ & $13.6$ & $4.6$\\
{\OVI}$_1$ & $-77$ & $-0.3$ & $-1.0$ & $20000$ & $9500$ & $0.0001$ & $18.6$ & $14.2$ & $9.4$ & $5.4$ & $14.2$ & $14.2$ & $28.2$ & $22.0$\\
{\OVI}$_2$ & $21$ & $-0.3$ & $-1.0$ & $20000$ & $19000$ & $0.0001$ & $18.9$ & $14.5$ & $9.7$ & $5.8$ & $14.5$ & $14.5$ & $34.7$ & $30.0$\\
{\OVI}$_3$ & $58$ & $-0.3$ & $-1.0$ & $20000$ & $4800$ & $0.0001$ & $18.3$ & $13.9$ & $9.1$ & $5.1$ & $13.9$ & $13.9$ & $26.7$ & $20.0$\\
{\Lya}$_1$ & $-137$ & $-3.5$ & $-2.0$ & $27000$ & $23000$ & $0.001$ & $20.0$ & $16.4$ & $9.8$ & $6.3$ & $12.1$ & $10.3$ & $15.0$ & ...\\


\hline
\enddata
\vglue -0.05in
\tablecomments{
\baselineskip=0.7\baselineskip
Model parameters for the system at $z=1.3250$.  Results of this model
are displayed superimposed on the data in Fig \ref{fig132}.
All column densities are expressed in units of {\cmsq}.  $b_{opt}$ is
the Doppler parameter for the transition that was ``optimized on'' in
modeling, the same transition as listed in the cloud designation column.  The ionization fraction, $f$, can be determinted from $\log f = \log {\rm N}({\HI}) - \log {\rm N}_{Tot}$(H).}
\label{val132}
\end{deluxetable}
\clearpage

\newpage
\begin{deluxetable}{cccccccccccccccccc}
\tablenum{8}
\tabletypesize{\scriptsize}
\rotate
\tablewidth{0pt}
\tablecaption{Model Parameters for $z=1.0480$}
\tablehead{
\colhead{Cloud}   &
\colhead{v}   &
\colhead{$\log{{\rm Z}\over{Z_{\sun}}}$}   &
\colhead{$\log~$U} &
\colhead{T} &
\colhead{Size} &
\colhead{n$_H$} &
\colhead{$\log$}   &
\colhead{$\log$}   &
\colhead{$\log$}   &
\colhead{$\log$}   &
\colhead{$\log$}   &
\colhead{$\log$}   &
\colhead{$\log$}   &
\colhead{b$_{\HI}$} &
\colhead{b$_{opt}$} \\
\colhead{}   &
\colhead{{\kms}}   &
\colhead{}   &
\colhead{} &
\colhead{K} &
\colhead{pc} &
\colhead{{\cc}} &
\colhead{N$_{Tot}$(H)}   &
\colhead{N({\HI})}   &
\colhead{N({\MgII})}   &
\colhead{N({\MgI})}   & 
\colhead{N({\SiIII})}   & 
\colhead{N({\SiIV})}   &
\colhead{N({\CIV})}   & 
\colhead{{\kms}}   &
\colhead{{\kms}}   
}

\startdata
{\MgII}$_1$ & $-97$ & $-0.7$ & $-3.8$ & $11000$ & $0.47$ & $0.061$ & $17.0$ & $15.5$ & $11.6$ & $9.6$ & $11.4$ & $9.4$ & $9.2$ & $13.4$ & $3.0$\\
{\MgII}$_2$ & $-86$ & $-0.7$ & $-4.1$ & $10000$ & $1.4$ & $0.12$ & $17.7$ & $16.5$ & $12.5$ & $10.5$ & $11.9$ & $9.6$ & $9.3$ & $13.7$ & $5.3$\\
{\MgII}$_3$ & $-23$ & $-0.7$ & $-4.0$ & $10000$ & $1.1$ & $0.097$ & $17.5$ & $16.2$ & $12.2$ & $10.2$ & $11.8$ & $9.5$ & $9.3$ & $16.7$ & $10.9$\\
{\MgII}$_4$ & $0$ & $-0.7$ & $-3.7$ & $11000$ & $9.9$ & $0.049$ & $18.2$ & $16.6$ & $12.8$ & $10.7$ & $12.7$ & $10.8$ & $10.7$ & $16.6$ & $10.3$\\
{\MgII}$_5$ & $6$ & $-0.7$ & $-3.7$ & $11000$ & $3.0$ & $0.049$ & $17.6$ & $16.0$ & $12.2$ & $10.2$ & $12.1$ & $10.2$ & $10.2$ & $13.8$ & $4.4$\\
{\MgII}$_6$ & $17$ & $-0.7$ & $-3.6$ & $11000$ & $6.3$ & $0.039$ & $17.9$ & $16.2$ & $12.4$ & $10.3$ & $12.4$ & $10.6$ & $10.6$ & $14.7$ & $6.4$\\
{\MgII}$_7$ & $34$ & $-0.7$ & $-4.0$ & $11000$ & $0.56$ & $0.097$ & $17.2$ & $15.9$ & $11.9$ & $10.0$ & $11.5$ & $9.3$ & $9.1$ & $15.7$ & $9.1$\\
{\SiIV}$_1$ & $6$ & $-1.4$ & $...$ & $55000$ & $200$ & $...$ & $19.6$ & $15.8$ & $10.3$ & $0.0$ & $13.6$ & $13.3$ & $12.3$ & $31.5$ & $11.0$\\
{\SiIII}$_1$ & $-186$ & $-1.4$ & $-2.1$ & $22000$ & $2000$ & $0.001$ & $18.9$ & $15.5$ & $11.1$ & $8.1$ & $12.7$ & $12.2$ & $13.1$ & $23.3$ & $14.0$\\
{\Lya}$_1$ & $-266$ & $-1.4$ & $-3.0$ & $15000$ & $1.3$ & $0.010$ & $16.6$ & $14.2$ & $9.8$ & $7.5$ & $10.6$ & $9.4$ & $9.8$ & $18.0$ & ...\\


\hline
\enddata
\vglue -0.05in
\tablecomments{
\baselineskip=0.7\baselineskip
Model parameters for the system at $z=1.0480$.  Results of this model
are displayed superimposed on the data in Fig \ref{fig104}.  A ``...''
entry in the {\IP} and $n_H$ columns indicates a collisionally ionized cloud.
All column densities are expressed in units of {\cmsq}.  $b_{opt}$ is
the Doppler parameter for the transition that was ``optimized on'' in
modeling, the same transition as listed in the cloud designation column.  The ionization fraction, $f$, can be determinted from $\log f = \log {\rm N}({\HI}) - \log {\rm N}_{Tot}$(H).}
\label{val104}
\end{deluxetable}
\clearpage

\newpage
\begin{deluxetable}{cccccccccccccccc}
\tablenum{9}
\tabletypesize{\scriptsize}
\rotate
\tablewidth{0pt}
\tablecaption{Model Parameters for $z=0.7290$}
\tablehead{
\colhead{Cloud}   &
\colhead{v}   &
\colhead{$\log{{\rm Z}\over{Z_{\sun}}}$}   &
\colhead{$\log~$U} &
\colhead{T} &
\colhead{Size} &
\colhead{n$_H$} &
\colhead{$\log$}   &
\colhead{$\log$}   &
\colhead{$\log$}   &
\colhead{$\log$}   &
\colhead{$\log$}   &
\colhead{$\log$}   &
\colhead{b$_{\HI}$} &
\colhead{b$_{opt}$} \\
\colhead{}   &
\colhead{{\kms}}   &
\colhead{}   &
\colhead{} &
\colhead{K} &
\colhead{pc} &
\colhead{{\cc}} &
\colhead{N$_{Tot}$(H)}   &
\colhead{N({\HI})}   &
\colhead{N({\MgII})}   & 
\colhead{N({\MgI})}   &
\colhead{N({\SiIV})}   &
\colhead{N({\CIV})}   & 
\colhead{{\kms}}   &
\colhead{{\kms}}   
}

\startdata
{\MgII}$_1$ & $-82$ & $0.7$ & $-3.0$ & $50$ & $0.97$ & $0.0059$ & $16.3$ & $15.4$ & $12.5$ & $10.0$ & $9.8$ & $11.9$ & $4.2$ & $4.1$\\
{\MgII}$_2$ & $-64$ & $0.7$ & $-4.1$ & $300$ & $0.029$ & $0.075$ & $15.8$ & $15.4$ & $12.1$ & $10.5$ & $8.4$ & $8.7$ & $4.1$ & $3.5$\\
{\MgII}$_3$ & $-9$ & $0.7$ & $-4.4$ & $70$ & $0.006$ & $0.15$ & $15.4$ & $15.2$ & $11.7$ & $10.6$ & $7.5$ & $8.1$ & $4.2$ & $4.1$\\
{\MgII}$_4$ & $12$ & $0.7$ & $-4.3$ & $300$ & $0.087$ & $0.12$ & $16.5$ & $16.1$ & $12.8$ & $11.3$ & $8.8$ & $9.0$ & $4.2$ & $3.6$\\
{\MgII}$_5$ & $67$ & $0.7$ & $-2.5$ & $600$ & $0.82$ & $0.0019$ & $15.7$ & $13.7$ & $11.5$ & $8.3$ & $10.8$ & $11.1$ & $5.8$ & $4.9$\\


\hline
\enddata
\vglue -0.05in
\tablecomments{
\baselineskip=0.7\baselineskip
Model parameters for the system at $z=0.7290$.  Results of this model
are displayed superimposed on the data in Fig \ref{fig072}.
All column densities are expressed in units of {\cmsq}.  $b_{opt}$ is
the Doppler parameter for the transition that was ``optimized on'' in
modeling, the same transition as listed in the cloud designation column.  The ionization fraction, $f$, can be determinted from $\log f = \log {\rm N}({\HI}) - \log {\rm N}_{Tot}$(H).}
\label{val072}
\end{deluxetable}
\clearpage

\newpage
\begin{deluxetable}{ccccccccccccccccc}
\tablenum{10}
\tabletypesize{\scriptsize}
\rotate
\tablewidth{0pt}
\tablecaption{Model Parameters for $z=0.5764$}
\tablehead{
\colhead{Cloud}   &
\colhead{v}   &
\colhead{$\log{{\rm Z}\over{Z_{\sun}}}$}   &
\colhead{$\log~$U} &
\colhead{T} &
\colhead{Size} &
\colhead{n$_H$} &
\colhead{$\log$}   &
\colhead{$\log$}   &
\colhead{$\log$}   &
\colhead{$\log$}   &
\colhead{$\log$}   &
\colhead{$\log$}   &
\colhead{b$_{\HI}$} &
\colhead{b$_{opt}$} \\
\colhead{}   &
\colhead{{\kms}}   &
\colhead{}   &
\colhead{} &
\colhead{K} &
\colhead{pc} &
\colhead{{\cc}} &
\colhead{N$_{Tot}$(H)}   &
\colhead{N({\HI})}   &
\colhead{N({\MgII})}   &
\colhead{N({\MgI})}   &
\colhead{N({\SiII})}   &
\colhead{N({\CIV})}   &
\colhead{{\kms}}   &
\colhead{{\kms}}   
}

\startdata
{\MgI}$_1$ & $-31$ & $-1.9$ & $-8.2$ & $7000$ & $0.00079$ & $700$ & $18.2$ & $18.2$ & $11.9$ & $10.6$ & $11.9$ & $0.0$ & $7.2$ & $1.8$\\
{\MgI}$_2$ & $-9$ & $-1.9$ & $-7.6$ & $1000$ & $0.031$ & $180$ & $19.2$ & $19.2$ & $12.9$ & $11.9$ & $12.9$ & $0.9$ & $11.2$ & $11.1$\\
{\MgI}$_3$ & $16$ & $-1.9$ & $-7.7$ & $6000$ & $0.018$ & $220$ & $19.1$ & $19.1$ & $12.8$ & $11.7$ & $12.8$ & $0.0$ & $20.9$ & $20.7$\\
{\SiII}$_1$ & $-31$ & $-1.9$ & $-3.0$ & $16000$ & $14000$ & $0.004$ & $20.3$ & $19.8$ & $13.5$ & $10.1$ & $14.0$ & $12.4$ & $20.6$ & $14.8$\\
{\SiII}$_2$ & $-9$ & $-1.9$ & $-2.5$ & $20000$ & $120000$ & $0.001$ & $20.8$ & $19.9$ & $13.6$ & $9.9$ & $13.8$ & $13.6$ & $21.1$ & $14.0$\\
{\SiII}$_3$ & $16$ & $-1.9$ & $-5.0$ & $10000$ & $100$ & $0.44$ & $20.1$ & $18.5$ & $13.8$ & $10.8$ & $13.8$ & $0.0$ & $17.7$ & $15.0$\\
{\MgII}$_1$ & $77$ & $-1.9$ & $-2.3$ & $22000$ & $22000$ & $0.001$ & $19.7$ & $16.4$ & $11.6$ & $8.5$ & $12.0$ & $13.0$ & $19.3$ & $5.0$\\
{\CIV}$_1$ & $-95$ & $-1.9$ & $-1.8$ & $30000$ & $140000$ & $0.0003$ & $20.1$ & $16.2$ & $11.2$ & $7.6$ & $11.8$ & $13.7$ & $23.0$ & $9.0$\\
{\CIV}$_2$ & $-3$ & $-1.9$ & $-1.8$ & $30000$ & $270000$ & $0.0003$ & $20.4$ & $16.6$ & $11.6$ & $7.8$ & $12.2$ & $14.0$ & $27.9$ & $18.8$\\
{\CIV}$_3$ & $20$ & $-1.9$ & $-1.8$ & $30000$ & $49000$ & $0.0003$ & $19.6$ & $15.8$ & $10.7$ & $7.3$ & $11.3$ & $13.3$ & $23.4$ & $10.0$\\
{\CIV}$_4$ & $37$ & $-1.9$ & $-1.8$ & $30000$ & $120000$ & $0.0003$ & $20.0$ & $16.2$ & $11.2$ & $7.5$ & $11.8$ & $13.7$ & $20.7$ & $6.0$\\
{\CIV}$_5$ & $54$ & $-1.9$ & $-1.8$ & $30000$ & $120000$ & $0.0003$ & $20.0$ & $16.2$ & $11.2$ & $7.5$ & $11.8$ & $13.7$ & $22.7$ & $8.0$\\


\hline
\enddata
\vglue -0.05in
\tablecomments{
\baselineskip=0.7\baselineskip
Model parameters for the system at $z=0.5764$.  Results of this model
are displayed superimposed on the data in Fig \ref{fig057}.  A ``...''
entry in the {\IP} column indicates a collisionally ionized cloud.
All column densities are expressed in units of {\cmsq}.  $b_{opt}$ is
the Doppler parameter for the transition that was ``optimized on'' in
modeling, the same transition as listed in the cloud designation column.  The ionization fraction, $f$, can be determinted from $\log f = \log {\rm N}({\HI}) - \log {\rm N}_{Tot}$(H).}

\label{val057}
\end{deluxetable}
\clearpage

\end{document}